\documentclass{gGAF2e}

\usepackage{hyperref}
\graphicspath{{./fig/}{./png/}}

\newcommand{\blue}{\textcolor{blue}}

\renewcommand{\blue}{\textcolor{black}}

\newcommand{\sgn}{{\rm sgn}}
\newcommand{\Ma}{{\rm Ma}}
\newcommand{\dd}{{\rm d}}
\newcommand{\ii}{{\rm i}}
\def\kNy{k_{\rm Ny}}

\def\xiM{\xi_{\rm M}}
\def\phiM{\phi_{\rm M}}
\def\EEK{{\cal E}_{\rm K}}
\def\EEM{{\cal E}_{\rm M}}
\def\EEGW{{\cal E}_{\rm GW}}

\def\EErad{{\cal E}_{\rm rad}}

\def\EEcrit{{\cal E}_{\rm crit}}

\def\EK{E_{\rm K}}
\def\EM{E_{\rm M}}

\def\hX{h_\times}
\def\hT{h_+}
\def\hX{h_\times}

\def\urms{u_{\rm rms}}
\def\hrms{h_{\rm rms}}

\def\kGW{k_{\rm GW}}

\def\la{\mathrel{\mathchoice {\vcenter{\offinterlineskip\halign{\hfil
$\displaystyle##$\hfil\cr<\cr\sim\cr}}}
{\vcenter{\offinterlineskip\halign{\hfil$\textstyle##$\hfil\cr<\cr\sim\cr}}}
{\vcenter{\offinterlineskip\halign{\hfil$\scriptstyle##$\hfil\cr<\cr\sim\cr}}}
{\vcenter{\offinterlineskip\halign{\hfil$\scriptscriptstyle##$\hfil\cr<\cr\sim\cr}}}}}
\def\ga{\mathrel{\mathchoice {\vcenter{\offinterlineskip\halign{\hfil
$\displaystyle##$\hfil\cr>\cr\sim\cr}}}
{\vcenter{\offinterlineskip\halign{\hfil$\textstyle##$\hfil\cr>\cr\sim\cr}}}
{\vcenter{\offinterlineskip\halign{\hfil$\scriptstyle##$\hfil\cr>\cr\sim\cr}}}
{\vcenter{\offinterlineskip\halign{\hfil$\scriptscriptstyle##$\hfil\cr>\cr\sim\cr}}}}}
\newcommand{\nab}{\bm{\nabla}}
\newcommand{\uu}{\bm{u}}
\newcommand{\ww}{\bm{w}}
\newcommand{\qq}{\bm{q}}
\newcommand{\QQ}{\bm{Q}}
\newcommand{\BB}{\bm{B}}
\newcommand{\JJ}{\bm{J}}
\newcommand{\ee}{\bm{e}}
\newcommand{\ff}{\bm{f}}
\newcommand{\kk}{\bm{k}}
\newcommand{\xx}{\bm{x}}
\newcommand{\SSSS}{\mbox{\boldmath ${\sf S}$} {}}
\newcommand{\EQ}{\begin{equation}}
\newcommand{\EN}{\end{equation}}
\newcommand{\EQA}{\begin{eqnarray}}
\newcommand{\ENA}{\end{eqnarray}}

\newcommand{\Eq}[1]{equation~(\ref{#1})}

\newcommand{\App}[1]{appendix~\ref{#1}}
\newcommand{\Sec}[1]{section~\ref{#1}}

\newcommand{\Fig}[1]{figure~\ref{#1}}
\newcommand{\FFig}[1]{figure~\ref{#1}}
\newcommand{\Figs}[2]{figures~\ref{#1} and \ref{#2}}

\newcommand{\Tab}[1]{table~\ref{#1}}

\newcommand{\bra}[1]{\langle #1\rangle}

\newcommand{\bbra}[1]{\left\langle #1\right\rangle}

\newcommand{\GeV}{\,{\rm GeV}}
\newcommand{\Mpc}{\,{\rm Mpc}}
\newcommand{\km}{\,{\rm km}}
\newcommand{\s}{\,{\rm s}}

\def\half{{\textstyle{1\over2}}}
\def\onethird{{\textstyle{1\over3}}}

\def\vA{v_{\rm A}}
\def\cs{c_{\rm s}}

\renewcommand{\Theta}{{\varTheta}}

\newcommand{\yan}[5]{, #5. {\em Astron.\ Nachr.\ } #1, {\bf #2}, #3--#4.}

\newcommand{\yanaN}[4]{, #4. {\em Astron.\ Astrophys.\ } #1, {\bf #2}, #3.}

\newcommand{\ysph}[5]{, #5. {\em Solar Phys.\ } #1, {\bf #2}, #3--#4.}

\newcommand{\ymn}[5]{, #5. {\em Month. Not. Roy.\ Astron.\ Soc.\ } #1, {\bf #2}, #3--#4.}

\newcommand{\yjfm}[5]{, #5. {\em J.\ Fluid Mech.\ } #1, {\bf #2}, #3--#4.}

\newcommand{\yprd}[5]{, #5. {\em Phys.\ Rev.\ D } #1, {\bf #2}, #3--#4.}

\newcommand{\yprdN}[4]{, #4. {\em Phys.\ Rev.\ D } #1, {\bf #2}, #3.}

\newcommand{\ypreN}[4]{, #4. {\em Phys.\ Rev.\ E } #1, {\bf #2}, #3.}

\newcommand{\sprl}[3]{, #2. {\em Phys.\ Rev.\ Lett.} #1, submitted, arXiv:#3.}
\newcommand{\yprl}[5]{, #5. {\em Phys.\ Rev.\ Lett.\ } #1, {\bf #2}, #3--#4.}
\newcommand{\yprlN}[4]{, #4. {\em Phys.\ Rev.\ Lett.\ } #1, {\bf #2}, #3.}

\newcommand{\yjcp}[5]{, #5. {\em J.\ Comp.\ Phys.\ } #1, {\bf #2}, #3--#4.}

\newcommand{\yapj}[5]{, #5. {\em Astrophys.\ J.\ } #1, {\bf #2}, #3--#4.}
\newcommand{\yapjN}[4]{, #4. {\em Astrophys.\ J.\ } #1, {\bf #2}, #3.}

\newcommand{\ycqgN}[4]{, #4. {\em Class.\ Quant.\ Grav.\ } #1, {\bf #2}, #3.}

\newcommand{\ypf}[5]{, #5. {\em Phys.\ Fluids } #1, {\bf #2}, #3--#4.}

\newcommand{\ypfN}[4]{, #4. {\em Phys.\ Fluids } #1, {\bf #2}, #3.}

\newcommand{\yjour}[6]{, #6. {\em #2} #1, {\bf #3}, #4--#5.}
\newcommand{\yjourN}[5]{, #5. {\em #2} #1, {\bf #3}, #4.}

\newcommand{\yproc}[7]{, #4. In {\em #5} (ed.\ #6) #1, pp.\ #2--#3.\ #7.}
\newcommand{\ybook}[3]{, #2. {\em #3} #1.}

\usepackage{verbatim}
\usepackage{xcolor}

\begin{document}

\jvol{114} \jnum{130} \jyear{2020} 

\markboth{\rm ROPER POL ET AL.}{\rm GEOPHYSICAL \& ASTROPHYSICAL FLUID DYNAMICS}
         


\title{The timestep constraint in solving the gravitational wave equations
sourced by hydromagnetic turbulence}

\author{Alberto Roper Pol$^{\rm a,b,c}$\thanks{$^\ast$Email: Alberto.RoperPol@colorado.edu},
Axel Brandenburg$^{\rm b,d,e,f}$, Tina Kahniashvili$^{\rm c,f,g}$, Arthur Kosowsky$^{\rm h}$
and Sayan Mandal$^{\rm f,c}$\vspace{6pt}
\\
$^{\rm a}$Department of Aerospace Engineering Sciences,
University of Colorado, Boulder, CO 80303, USA
\\
$^{\rm b}$Laboratory for Atmospheric and Space Physics,
University of Colorado, Boulder, CO 80303, USA
\\
$^{\rm c}$Abastumani Astrophysical Observatory, Ilia State University,
3-5 Cholokashvili St., 0194 Tbilisi, Georgia
\\
$^{\rm d}$JILA and Department of Astrophysical and Planetary Sciences, University of Colorado,\\
Boulder, CO 80303, USA
\\
$^{\rm e}$NORDITA, KTH Royal Institute of Technology and Stockholm University, and\\
Department of Astronomy, Stockholm University, SE-10691 Stockholm, Sweden
\\
$^{\rm f}$McWilliams Center for Cosmology and Department of Physics, Carnegie Mellon University,\\
5000 Forbes Ave, Pittsburgh, PA 15213, USA
\\
$^{\rm g}$Department of Physics, Laurentian University, Ramsey
Lake Road, Sudbury, ON P3E 2C, Canada
\\
$^{\rm h}$Department of Physics and Astronomy, University of Pittsburgh, and
Pittsburgh Particle Physics, Astrophysics, and Cosmology Center (PITT PACC), Pittsburgh, PA 15260, USA
\vspace{6pt}\received{\it \today,~ $ $Revision: 1.310 $ $} }

\maketitle

\begin{abstract}
Hydromagnetic turbulence produced during phase transitions
in the early universe can be a powerful source of stochastic gravitational
waves (GWs).
GWs can be modelled by the linearised spatial part of the Einstein
equations sourced by the Reynolds and Maxwell stresses.
We have implemented two different GW solvers into the {\sc Pencil
Code}---a code which uses a third order timestep and sixth order finite
differences.
Using direct numerical integration of the GW equations, we study the
appearance of a numerical degradation of the GW amplitude at the highest
wavenumbers, which depends on the length of the timestep---even when the
Courant--Friedrichs--Lewy condition is ten times below the stability limit.
This degradation leads to a numerical error, which is found to scale
with the third power of the timestep.
A similar degradation is not seen in the magnetic and velocity fields.
To mitigate numerical degradation effects, we alternatively use the
exact solution of the GW equations under the assumption that the source
is constant between subsequent timesteps.
This allows us to use a much longer timestep, which cuts the
computational cost by a factor of about ten.

\end{abstract}

\section{Introduction}

Wave equations coupled to fluid equations appear in at least two
different contexts.
The Lighthill equation in aeroacoustics is one such example
\citep{Lig52,Lig54}, and the linearised  gravitational wave (GW) equation
is another \citep[e.g.,][]{Gri74,Deryagin:1986qq}.
The former example is important not only in aviation, where the Lighthill
equation is used to quantify the sound production from jet engines, but
it is also relevant to stars with outer convection zones, where sound
waves from the outer layers can be responsible for chromospheric and
coronal heating \citep{Ste67}.
Stochastic GWs, on the other hand, are expected to
be generated in the early universe by hydrodynamic
turbulence, as discussed in early papers by \cite{KKT94} and \cite{KMK02}.
GWs are also expected from magnetohydrodynamic (MHD) turbulence; see
\cite{DFK00}.
Recently, the GW signal from MHD turbulence has been studied in \cite{NSS18}
and \cite{STY18} \citep[see also][for a review, and references therein]{CF18}.
It is therefore of interest to solve the MHD
equations simultaneously with the GW equation.
Both the MHD and GW equations
are three-dimensional partial differential equations that can be
solved with similar numerical techniques.
However, the numerical properties are not quite
equivalent and the physical intuition gained from numerical
hydrodynamics gives insufficient guidance on the numerical requirements
for the length of the timestep.

For the numerical solution of the fluid equations, often 
the accuracy of the solution is not strongly affected by the timestep.
Therefore, in practice, one is able to use a timestep that is close
to the stability limit of the scheme.
However, with a finer timestep, more steps are
needed to cover a given time span, so increased error accumulation
is a possibility.
The situation seems to be different for the solution of a wave equation
sourced by hydrodynamic and magnetic stresses.
An accurate representation of the high wavenumber
contributions hinges sensitively on the length of the timestep adopted.
This leads to an artificial drop of the GW
spectral energy density at high wavenumbers, due to the inaccuracy
in the solution, if the timestep is not small enough.

In numerical hydrodynamics and MHD, the maximum permissible timestep
$\delta t$ is given by the Courant--Friedrichs--Lewy (CFL) condition
\citep{CFL28},
\EQ
\delta t\,\leq\, C_{\rm CFL} \delta x/U_{\rm eff}\,,
\EN
where $C_{\rm CFL}$ is a number of the order of unity, $\delta x$ is
the mesh width, and $U_{\rm eff}$ is an effective propagation speed.
This could be the advection speed $\uu$, the sound speed $\cs$,
the Alfv\'en speed $\vA$ in the presence of magnetic fields, or,
more generally, a combination of various relevant speeds such as
$|\uu|+(\cs^2+\vA^2)^{1/2}$, which is the expression for the
Doppler-shifted fast magnetosonic wave speed.
The CFL condition is a necessary condition for the
stability
of explicit time integration and upwind schemes
in hyperbolic equations (e.g., wave or convection equations),
where information travels a distance
$U_{\rm eff} \delta t$ within one time step $\delta t$
\citep[see chapter~8.3 in][]{Ferziger}.
In addition, the CFL condition also affects more complex partial
differential equations, which include the presence of waves. This is the case for
the MHD equations.
Under these circumstances, the CFL condition is an approximation to the required
condition for stability, where the exact value of $C_{\rm CFL}$
depends on the time stepping scheme.

A major difference between the hydrodynamic and the aforementioned
aeroacoustic and GW equations lies in the fact that the former are nonlinear.
Therefore, even though there may also be an inverse cascade, some energy
always cascades down to smaller length scales.
This does not happen with linear wave equations with constant coefficients.
Here, instead, the high wavenumber contributions are only excited because
they are being sourced at those high wavenumbers.

\section{Basic equations}

\subsection{Aeroacoustic and GW applications}

The aeroacoustic or Lighthill equation
describes the generation of acoustic waves propagating away from
a turbulent source in a stationary medium outside the region of turbulence
\citep{GD17}.
It can be written in the form due to
\citep{Lig52,Lig54}
\EQ
\left({\upartial^2\over\upartial t^2}-\cs^2\nabla^2\right)
\delta\rho (\xx, t)\,=\,{\upartial^2 T_{ij}(\xx, t)
\over\upartial x_i\upartial x_j}\,,
\label{Lighthill_eqn}
\EN
where $\cs$ is the sound speed of the stationary background medium,
which is constant,
$\delta\rho$ is the density fluctuation in the stationary medium,
$t$ and $\xx$, in the context of this equation, to physical space and time coordinates
(as opposed to comoving space and conformal time coordinates used in the GW
equation discussed below), and $T_{ij}(\xx, t) = \rho u_i u_j$ is
the stress tensor with $\rho$ being the
mass density and $\uu$ the turbulent velocity.
It has been applied to sound generation from isotropic
homogeneous turbulence by \cite{Pro52}.
In this work, it was found that the efficiency of sound production, i.e., the
ratio between acoustic power and the rate of kinetic energy dissipation,
$\epsilon_{\rm K}\approx\urms^3/\ell$, scales with the Mach number $\Ma=\urms/\cs$
to the fifth power, where $\urms$ is the root mean square velocity
perturbation and $\ell$ is the turbulence length scale.
The intensity of sound waves is related to the fluctuation of the pressure field,
which, in turn, is related to the density perturbations through the sound speed $\cs$
as $\delta p (\xx, t)=\cs^2 \, \delta\rho (\xx, t)$.
The acoustic intensity in a statistically stationary homentropic medium,
which holds far from the
turbulent source, is defined as the expectation value of $\delta p\, \uu$. In Fourier space,
using hydrodynamic momentum conservation, one obtains the radial
intensity, i.e., the intensity of the sound wave propagating away from the source,
as $\tilde{I}_{\rm r} (\kk, t) =
| \widetilde{\delta} p (\kk, t)|^2/(2 \rho \cs)$,
where $\widetilde{\delta p} (\kk, t)$ is the Fourier amplitude of the
pressure perturbation $\delta p (\xx, t)$ \citep{GD17}.

The direct analogy between sound generation and GW
generation from isotropic homogeneous turbulence was exploited
by \cite{Gogo07} and \cite{KGR08}.
The aeroacoustic analogy allows approximated analytical description of turbulence
generated GWs. In this work we focus on the numerical aspects of GW and MHD
equations, and the generation of sound waves, as well as the aeroacoustic analogy
for GW generation, is left for future studies.
In a flat expanding universe, during the radiation-dominated epoch,
using comoving spatial coordinates
(which follow the expansion of the universe) and conformal time
$t$ (related to physical time $t_{\rm phys}$ through
$\dd t_{\rm phys}=a(t) \dd t$),
the GW equation (see appendix~\ref{normalised_eqn} for details) is
given by (\ref{GW3}).
\EQ
\left({\upartial^2\over\upartial t^2}-c^2\nabla^2\right)
h_{ij}^{\rm TT} (\xx, t)\, =\, {16\pi G\over a(t)
c^2} T_{ij}^{\rm TT} (\xx, t)\,,
\label{GW_eqn}
\EN
where ${\bm \nabla}$ refers to comoving spatial derivatives,
$a (t)$ is the scale factor in the
Friedmann--Lema\^itre--Robertson--Walker (FLRW) model,
which exhaustively describes the metric tensor
in a spatially flat isotropic and homogeneous universe,
$h_{ij}^{\rm TT} (\xx, t)$ are the scaled tensor-mode perturbations of the metric tensor,
also called (scaled) strains, related to physical strains
through $h_{ij}^{\rm TT}(\xx, t) = a(t) h_{ij, \text{phys}}^{\rm TT} (\xx, t)$,
such that the spatial components of the metric tensor are
$g_{ij} (\xx, t) = a^2 (t) [ \delta_{ij} +
h_{ij, \text{phys}}^{\rm TT} (\xx, t) ]$,
$\delta_{ij}$ being the Kronecker delta;
$c$ is the speed of light,
$G$ is Newton's gravitational constant,
and $T_{ij}^{\rm TT}(\xx, t)$ is
the transverse and traceless (TT) projection of the comoving stress--energy
tensor $T_{ij} (\xx, t)$ (see \citealt{Gri74} and \citealt{Deryagin:1986qq}).
Since (\ref{GW_eqn}) is the result of linearisation in unbounded space,
we assume that the spatial average of $T_{ij}^{\rm TT}(\xx, t)$ vanishes.
During the radiation-dominated epoch, $a (t)$ evolves
linearly with conformal time, as inferred from the Friedmann equations
\citep{Friedmann1922} for a perfect fluid with relativistic equation of
state $p = \rho c^2/3$.
Hence, in (\ref{GW_eqn}) there is no damping term due to the
expansion of the universe; see appendix~\ref{normalised_eqn} for details.

The TT projection \citep[see box 35.1 of][]{MTW73} can be computed
in Fourier space (indicated by a tilde) as
\EQ
\tilde{T}_{ij}^{\rm TT}(\kk,t)\,=\,\bigl(P_{il}P_{jm}-\half P_{ij}P_{lm}\bigr)\,
\tilde{T}_{lm}(\kk,t)\hskip 8mm \mbox{for $|\kk|>0$}\,,
\label{Tij_TT}
\EN
where $P_{ij}(\kk)=\delta_{ij}-\hat k_i\hat k_j$ is the projection
operator, $\kk$ is the wavevector, and $\hat{\kk}=\kk/k$ is its unit
vector with $k=|\kk|$ being the modulus.
The stress-energy tensor $T_{ij}$ consists of the sum of negative
Reynolds and Maxwell stresses,
expressed for flat space-time geometry as
\EQ
T_{ij}\, = \,(p+\rho c^2)\gamma^2 u_i u_j/c^2 + p\delta_{ij}
-B_i B_j/\mu_0+\delta_{ij}\BB^2/(2\mu_0)+\cdots\,,
\label{Tij}
\EN
where $p$ is the fluid pressure,
$\EErad = \,\rho c^2$ is the radiation energy density,
$\rho$ being a normalised energy density (not to be confused with the
previously introduced mass density),
$\gamma=(1-\uu^2/c^2)^{-1/2}$ is the Lorentz factor,
$\uu$ is the turbulent velocity,
$\BB$ is the magnetic field,
$\mu_0$ is the vacuum permeability and the ellipsis denotes viscous
and resistive contributions that are ignored here.
From now on, we adopt Lorentz--Heaviside units for the magnetic fields,
such that $\mu_0 = 1$.
The MHD fields ($p$, $\rho$, $\uu$ and $\BB$) that appear in (\ref{Tij})
are functions of comoving space and conformal time coordinates.
They are expressed as comoving fields, leading to the comoving
stress-energy tensor, as described in \App{normalised_eqn}.

The GW \Eq{GW_eqn} can be rewritten
using normalised comoving space $\bar{\xx} =  \xx \, c/H_*$, and conformal time
$\bar{t} = t/t_*$, normalised Laplacian
$\bar\nabla^2 = H_*^2 \nabla^2/c^2$, and normalised stress-energy tensor
$\bar{T}_{ij}^{\text{TT}} (\bar{\xx}, \bar{t}) = T^{\text{TT}}_{ij}
(\xx, t)/\EErad^*$,
where $t_*$, $H_*$, and $\EErad^*$
are time, the Hubble parameter $H (t) = \dot{a}/a =a'/a^2$
(with the dot denoting derivative with respect to physical time $t_{\rm phys}$,
and the prime denoting derivative with respect to normalised conformal time $t$),
and the radiation energy density; see (\ref{rho_rad}),
respectively, all taken during the time of generation
(see \App{normalised_eqn} for details).
The normalised equation during the radiation-dominated epoch is
\begin{equation}
\left( \frac{\upartial^2}{\upartial t^2} - \nabla^2 \right)
h_{ij}^{\text{TT}} (\xx, t)\, =\, \frac{6}{t} T_{ij}^{\text{TT}} (\xx, t)\,,
\label{normalised_GW}
\end{equation}
where the overbar has been omitted from (\ref{GW4}), so we will be referring to
normalised comoving space and conformal time coordinates, Laplacian operator and
stress--energy tensor, as well as
normalised wavenumber and frequency, from now on
unless otherwise stated.

\subsection{TT projection and linear polarisation basis}
\label{TT}

The six components of the spatial part of the symmetric tensor
$h_{ij} (\xx, t)$, characterising the linearised evolution of the scaled strains,
contain four degrees of gauge freedom.
In the TT gauge, these are eliminated by requiring
$\tilde{h}_{ij}^{\rm TT} (\kk, t)$ to be a transverse and traceless tensor, i.e.,
$\tilde{h}_{ii}^{\rm TT} (\kk, t)=0$, and $k_j \tilde{h}_{ij}^{\rm TT} (\kk, t) = 0$,
respectively, where Einstein summation convention is being used,
leaving only two independent components which, in the
linear polarisation basis, are the $+$ and $\times$ polarisation modes.
To compute the physically observable characteristic amplitude,
GW energy density, and the degree of circular polarisation,
we compute $\tilde{h}_{ij}^{\rm TT} (\kk, t)$ and
$\tilde{h}_{ij}^{' \rm TT} (\kk, t)$, and express
them in terms of the linear polarisation modes.
The observable spectra of interest are defined and derived in more detail
in \Sec{GW_energy_density}.
To compute $\tilde{h}_{ij}^{\rm TT} (\kk, t)$ from $h_{ij}
(\xx, t)$, we take the Fourier transform of the six components of $h_{ij}
(\xx, t)$ using the convention
\EQ
\tilde{h}_{ij}(\kk,t)\,=\,\int h_{ij}(\xx,t)\,e^{-\ii\kk\cdot\xx}\, \dd^3\xx\,,
\EN
for $1\leq i\leq j\leq3$ and compute the components in the TT gauge as
\EQ
\tilde{h}_{ij}^{\rm TT}(\kk,t)\,=\,(P_{il}P_{jm}-\half P_{ij}P_{lm})\,
\tilde{h}_{lm}(\kk,t)\,.
\EN
Next, we compute the linear polarisation basis,
\begin{eqnarray}
e^+_{ij} (\kk)\,\,=\,e_i^1 e_j^1 - e_i^2 e_j^2\,,\hskip 10mm
e^\times_{ij} (\kk) \,=\,e_i^1 e_j^2 + e_i^2 e_j^1\,,
\end{eqnarray}
where $\ee^1$ and $\ee^2$ are unit vectors perpendicular to $\kk$
and perpendicular to each other.
This polarisation basis has the following orthogonality property
\begin{equation}
e^{+}_{ij} (\kk) e^{+}_{ij} (\kk)\, =\, e^{\times}_{ij} (\kk) e^{\times}_{ij}
(\kk)\, =\, 2\,, \hskip 10mm e^{+}_{ij} (\kk) e^{\times}_{ij} (\kk) \,=\, 0\,.
\label{orthogonality}
\end{equation}
Thus, the strains are decomposed into the two independent $+$ and $\times$
modes, such that $\tilde{h}_{ij}
^{\rm TT}(\kk, t) = e^+_{ij} (\kk) \tilde{h}_+(\kk, t) +
e^\times_{ij} (\kk) \tilde{h}_\times(\kk, t)$, with
\begin{equation}
\tilde{h}_+(\kk,t)\,=\,\half e^+_{ij}(\kk) \, \tilde{h}^{\rm TT}_{ij}(\kk,t)\,,
\hskip 10mm \tilde{h}_\times(\kk,t) \,=\, \half e^\times_{ij}(\kk) \,
\tilde{h}^{\rm TT}_{ij}(\kk,t)\,.
\end{equation}
We then return into physical space and compute
\EQ
h_{+, \times}(\xx,t)\,=\,{1 \over (2\pi)^3} \int \tilde{h}_{+, \times}(\kk,t)\,e^{\ii\kk\cdot\xx}
\, \dd^3\kk\,.
\EN
Analogous calculations are performed to compute $\tilde{h}'_{+, \times}(\kk,t)$
and $\tilde{T}_{+, \times} (\kk, t)$.
The normalised GW \Eq{normalised_GW} can be expressed for the two
independent $+, \times$ modes, in Fourier space, as
\begin{equation}
\left( \frac{\upartial^2}{\upartial t^2} + \kk^2 \right) \tilde{h}_{+, \times} (\kk, t)\, =\,
\frac{6}{t} \tilde{T}_{+, \times} (\kk, t)\,.
\end{equation}

\subsection{Choice of unit vectors in Fourier space}
\label{SignSwap}

The linear polarisation basis is formed by $\ee^1$,
$\ee^2$, and $\ee^3 = \hat{\kk}$,
for all wavevectors except $\pmb{k} = \ \pmb{0}$,
that corresponds to a monochromatic uniform field, and
is neglected in GW spectra because uniform fields do not generate GWs.
To construct $\ee^1$ and $\ee^2$ from $\kk$, we distinguish three cases
\begin{align}
\mbox{for}\;|k_1|\,<\,\min(|k_2|,|k_3|)&:&&&&\nonumber\\
\ee^1\,=\,\sgn&(\kk)\,(0,-\hat{k}_3,\hat{k}_2)\,, &
\ee^2\,=\,&\,(\hat{k}_2^2+\hat{k}_3^2,-\hat{k}_1\hat{k}_2,-\hat{k}_1 \hat{k}_3)\,,
\label{eek1}\\
\mbox{for}\;|k_2|\,<\,\min(|k_3|,|k_1|)&: &&&&\nonumber\\
\ee^1\,=\,\sgn&(\kk)\,(\hat{k}_3,0,-\hat{k}_1)\,,&
\ee^2\,=\,&\,(-\hat{k}_2\hat{k}_1,\hat{k}_3^2+\hat{k}_1^2,-\hat{k}_2 \hat{k}_3)\,,
\label{eek2}\\
\mbox{for}\;|k_3|\leq\, \min(|k_1|,|k_2|)&: &&&&\nonumber\\
\ee^1\,=\,\sgn&(\kk)\,(-\hat{k}_2,\hat{k}_1,0)\,, &
\ee^2\,=&\,\,(-\hat{k}_3\hat{k}_1,-\hat{k}_3\hat{k}_2,\hat{k}_1^2+\hat{k}_2^2)\,,
\label{eek3}
\end{align}
where we define the sign of a general wavevector $\kk = (k_1, k_2, k_3)$
in the following way
\begin{equation}
\sgn (\kk) = \left\{\begin{array}{l l}
\sgn (k_3) & \hskip 3mm \text{if }\hskip 2mm k_3 \neq 0\,,  \\[0.2em]
\sgn (k_2) & \hskip 3mm \text{if }\hskip 2mm k_3 = 0 \text{ \ and \ } \ k_2 \neq 0\,,  \\[0.2em]
\sgn (k_1) & \hskip 3mm \text{if }\hskip 2mm k_2 = k_3 = 0\,,
\end{array} \right.
\end{equation}
such that half of the wavevectors are considered positive and the other
corresponding half of the wavevectors are considered negative.
The way to choose which half of the wavevectors are positive is arbitrary
and could be changed leading to the same final result.

Note that neither $\ee^1$ nor $\ee^2$ flip sign under the parity
transformation $\kk\to-\kk$.
The reason for the $\sgn (\kk)$ term is the following.
The linear polarisation tensorial basis $e_{ij}^+ (\kk)$
and $e_{ij}^\times (\kk)$
must be represented by even operators with respect
to $\kk$ to reproduce the required modes,
as will be shown in next section with a simple example, a 
one-dimensional Beltrami field.
Alternatively, without loss of generality, we could have defined
$\ee^1$ and $\ee^2$ such that both flip sign under $\kk
\rightarrow - \kk$ transformations, such that both $e_{ij}^+
(\kk)$ and $e_{ij}^\times (\kk)$ tensors are even operators.

\subsection{Characteristic amplitude and GW energy density}
\label{GW_energy_density}
The most important physical observables related to stochastic primordial
GWs are the characteristic amplitude $\hrms(t)$ and the normalised
GW energy density $\varOmega_{\rm GW} (t)$. In this section, we give a
definition of these functions in terms of the physical strains
$h_{ij, {\rm phys}}^{\rm TT} (\xx, t)$.
We also define the spectral functions $\hrms (k, t)$,
$\varOmega_{\rm GW} (k, t)$, and $\varXi_{\rm GW} (k, t)$, that are useful to
characterise the energy and polarisation of GWs.
The characteristic amplitude $\hrms (t)$ of GWs is defined as
\begin{equation}
h^2_{\rm rms} (t) \,= \,\tfrac{1}{2} \bigl\langle h_{ij, {\rm phys}}^{\rm TT}
(\xx, t)  h_{ij, {\rm phys}}^{\rm TT} (\xx, t)\bigr\rangle\,,
\label{hrms_def}
\end{equation}
where angle brackets indicate volume averaging in physical space.
The mean GW energy density $\EEGW(t)$ is computed from the physical
time derivative of the physical strains $\dot{h}_{ij, {\rm phys}}^{\rm TT} (\xx, t)$ as
\begin{gather}
\EEGW (t)\, = \, \frac{c^2}{32 \pi G} \bbra{\dot{h}_{ij, {\rm phys}}^{\rm TT} (\xx, t)
\dot{h}_{ij, {\rm phys}}^{\rm TT} (\xx, t)}.
\label{EEGW_def}
\end{gather}
The normalised GW energy density is
$\varOmega_{\rm GW} (t) = \EEGW (t)\left/\EErad^*\right.$,
where $\EErad^*$ is the radiation energy density, defined in (\ref{rho_rad}).
The details on the computation of these quantities from the numerical code are
given in \App{spectral}.

The influence of velocity and magnetic fields
on the sourcing stress-energy tensor $T_{ij} (\xx, t)$
depends on the kinetic and magnetic energies.
These are defined from the physical density, velocity and magnetic fields as
$\EEK(t)=\bra{\rho(\xx, t)\uu^2(\xx, t)}/2$ and $\EEM(t)=
\bra{\BB^2(\xx, t)}/2$.
In analogy to the GW energy density, the
normalised magnetic and kinetic energy densities are
defined as $\varOmega_{\rm M, K} (t) = {\cal E}_{\rm M, K} (t) \left/ \EErad^* \right.$.
It is common notation to use $\varOmega$ for
normalised energy densities in cosmology.

We are interested in the spectra related to the physical observables described above,
in terms of their dependence on wavenumber $k$ and, hence,
frequency $f$, related to each other through the dispersion relation for GWs,
$2 \pi f = k$.
Note that $k$ and $f$ are comoving and normalised as described in
\App{normalised_eqn}.
The spectral function associated with the characteristic amplitude is
$h^2_{\rm rms}(k, t) = k S_h(k, t)$, where $S_h$ is defined in \App{spectral};
see (\ref{Sh_app}), such that
the characteristic amplitude $\hrms (t)$ is
\begin{equation}
h_{\rm rms}^2 (t) \,= \,\int_{-\infty}^{\infty} h_{\rm rms}^2 (k, t) \, \dd \ln k\,.
\end{equation}
Note that $\hrms(t)$ and its spectral function $\hrms(k, t)$ are distinguished
by the explicit specification of the argument $k$. We use the same symbol
for both since they represent the same physical quantity.

The corresponding spectral function for the normalised energy spectrum is
\begin{equation}
\varOmega_{\rm GW} (k, t) \,=\, \frac{k S_{\dot{h}} (k, t)}{6H_*^2}\,, \hskip 8mm
\text{such that \ }\hskip 4mm \int_{-\infty}^{\infty} \varOmega_{\rm GW} (k, t) \, \dd  \ln k\, =\,
\varOmega_{\rm GW} (t)\,,\hskip 8mm
\label{OmegaGW_app}
\end{equation}
where $S_{\dot{h}} (k, t)$ is defined in \App{spectral}; see (\ref{Sdoth_app}),
and $\varOmega_{\rm GW} (t)$ corresponds to the quantity defined in (\ref{EEGW_def}).
Note that both $\hrms (k, t)$, and $\varOmega_{\rm GW} (k, t)$ are
the spectra
defined per logarithmic wavenumber interval.
In analogy to the normalised energy spectrum $\varOmega_{\rm GW} (k, t)$,
the antisymmetric or helical energy spectrum $\varXi_{\rm GW} (k, t)$ is defined as
\begin{equation}
\varXi_{\rm GW}(k, t) \,=\, \frac{k A_{\dot{h}} (k, t)}{6 H_*^2}, \hskip 8mm \text{such that \ }\hskip 4mm
\int_{-\infty}^{\infty} \varXi_{\rm GW}(k, t) \, \dd \ln k\, =\, \varXi_{\rm GW}(t)\,, \hskip 8mm
\label{XiGW_app}
\end{equation}
where $A_{\dot{h}} (k, t)$ is defined in \App{spectral}; see (\ref{Adoth_app}), and
$\varXi_{\rm GW}(t)$ is the total
normalised helical energy density. The symmetric, $S_{\dot{h}}$,
and antisymmetric, $A_{\dot{h}}$, spectral functions, used to
characterise the tensorial field,
$\dot{h}_{ij, {\rm phys}}^{\rm TT}$, are described in detail in \App{spectral}.
The polarisation degree of GWs is defined as the ratio between the
antisymmetric and symmetric spectra.
\begin{equation}
\mathcal{P} (k, t) \,=\, \frac{\varXi_{\rm GW} (k, t)}
{\varOmega_{\rm GW} (k, t)}\, = \,\frac{A_{\dot{h}} (k, t)}{S_{\dot{h}} (k, t)}\,.
\end{equation}
Again, in analogy to the GW spectra, we define normalised
magnetic and kinetic spectra as
\begin{equation}
\varOmega_{\rm M, K} (k, t)\, =\, \frac{k E_{\rm M, K} (k, t)}{\EErad^*},\hskip 8mm
\text{such that \ }\hskip 4mm
\int_{-\infty}^{\infty} \varOmega_{\rm M, K} (k, t) \, \dd \ln k\, =\,
\varOmega_{\rm M, K} (t)\,,\hskip 8mm
\end{equation}
where the magnetic and kinetic energy spectra $E_{\rm M, K} (k, t)$ are defined
in \App{spectral}; see (\ref{EM}) and (\ref{EK}), and the total $\varOmega_{\rm M, K} (t)$ corresponds
to the normalised magnetic and kinetic energy densities, defined above.
Note that these spectra are defined per logarithmic wavenumber interval, just like
the normalised GW energy spectrum $\varOmega_{\rm GW} (k, t)$.

\subsection{Hydromagnetic equations}
\label{hydromagnetic}

There is a striking analogy between the normalised radiation energy
density $\rho = \EErad/c^2$ in the present context of
an ultrarelativistic plasma and
the mass density in the usual MHD equations
and in the Lighthill equation (\ref{Lighthill_eqn}).
Note that here ``normalised'' just refers to
division by $c^2$, not the
normalisation presented in \App{normalised_eqn}.
This analogy was employed by \cite{CHB01} and \cite{BJ04} to argue
that the equations for the early universe could be solved using just
ordinary MHD codes.
Here and below, we expand $\gamma$ in $\uu/c$, i.e.,
$\gamma \sim 1 + \frac{1}{2} \uu^2/c^2$,
including only terms up to second order.
The $p/c^2$ term enters in the stress--energy tensor
because in a relativistic plasma, the gas pressure
is equal to one third of the radiation energy
density, $\rho c^2$.
Using the ultrarelativistic equation of state, we have $p=\rho c^2/3$,
so the prefactor of $\gamma^2 u_i u_j$ in (\ref{Tij}) reduces to $4\rho/3$.
Hence, similar 4/3 factors appear in the MHD equations for an
ultrarelativistic gas in a flat expanding universe
\citep{BEO96,BKMRPTV17,KBDTY17}, which are given by
\begin{align}
{\upartial\ln\rho\over\upartial t}\,=\,&\,-\,\frac{4}{3}\left(\nab{\bm \cdot}\uu+\uu{\bm \cdot}\nab\ln\rho\right)
+{1\over\rho c^2}\left[\uu{\bm \cdot}(\JJ\times\BB)+\eta\JJ^2\right],
\label{dlnrhodt}\\
{\upartial\uu\over\upartial t}\,=\,&\,-\,\uu{\bm \cdot}\nab\uu
+{\uu\over3}\left(\nab{\bm \cdot}\uu+\uu{\bm \cdot}\nab\ln\rho\right)
-{\uu\over\rho c^2}\left[\uu{\bm \cdot}(\JJ\times\BB)+\eta\JJ^2\right] \nonumber \\
&\,-{c^2\over4}\nab\ln\rho
+{3\over4\rho}\JJ\times\BB+{2\over\rho}\nab{\bm \cdot}\left(\rho\nu\SSSS\right),
\label{dudt} \\
{\upartial\BB\over\upartial t}\,=\,&\,{\bm \nabla}\times(\uu\times\BB-\eta\JJ)\,,
\label{dAdt}
\end{align}
where ${\sf S}_{ij}=\half(u_{i,j}+u_{j,i})-\onethird\delta_{ij}\nab{\bm \cdot}\uu$
are the components of the rate-of-strain tensor with commas denoting partial
derivatives, $\JJ$ is the current density, $\nu$ is the viscosity,
and $\eta$ is the magnetic diffusivity.
We assume constant $\eta$ and $\nu$ in all our simulations.
We emphasise that all variables have been scaled appropriately so that
terms proportional to $\dot{a}/a$
appear neither in (\ref{dlnrhodt})--(\ref{dAdt})
nor in (\ref{GW_eqn}), i.e., we use comoving variables that already take into
account the effect of the expansion of the universe; see \App{normalised_eqn}.
Nevertheless, there remains the $a^{-1} (t)$ term on the right-hand side of
the GW \Eq{GW_eqn}, which means that the source of GWs
gradually declines during the radiation-dominated epoch of the universe.

Returning to the proposal of \cite{CHB01} and \cite{BJ04} to use just
ordinary MHD computer codes for simulating the early universe,
it is useful to compare with the non-relativistic limit.
Assuming $p \ + \BB^2/2\ll\rho c^2/3$, (\ref{dlnrhodt}) and (\ref{dudt}) reduce to
\begin{align}
{\upartial\ln\rho\over\upartial t}\,=\,&\,-\,\left( \nab{\bm \cdot}\uu
+\uu {\bm \cdot} \nab \ln \rho \right),
\label{dlnrhodt_nonrel} \\
{\upartial\uu\over\upartial t}\,=\,&\,-\,\uu{\bm \cdot}\nab\uu
-{1\over \rho}\nab p
+{1\over\rho}\JJ\times\BB+{2\over\rho}\nab{\bm \cdot}\left(\rho\nu\SSSS\right),
\label{dudt_nonrel}
\end{align}
amended by an isothermal equation of state with $p=\rho\cs^2$
\citep{CHB01,BJ04}, and (\ref{dAdt}) is unchanged.
However, in this paper we do not use the simplified equations even though
the difference in the final results between the two sets of equations is small: the
kinetic energy is less by a factor $4/3$ in the relativistic case in the
magnetically dominated case, while in the magnetically subdominant case,
the magnetic energy is larger by a factor $4/3$ \citep[see][]{BKMRPTV17}.

\subsection{Two approaches to solving the GW equation}
\label{two_approaches}

The decomposition into the linear polarisation basis described 
in \Sec{TT} is applied to the normalised GW \Eq{normalised_GW}.
However, the projection of $\, \tilde{T}_{ij}^{\rm TT} (\kk, t)$
onto $\tilde{T}_{+, \times} (\kk, t)$ at every timestep of the numerical simulation,
is computationally expensive,
because it requires nonlocal operations involving Fourier transformations.
When numerically integrating the GW equations, it is
therefore advantageous to instead evolve the scaled strains,
$h_{ij} (\xx, t)$, in an arbitrary gauge (computed from the
GW equation sourced by the unprojected $T_{ij} (\xx, t)$, instead of
$T_{ij}^{\rm TT} (\xx, t)$),
compute $\tilde{h}_{ij}^{\rm TT} (\kk, t)$,
and then perform the decomposition into the linear polarisation
basis whenever we compute physical quantities such as averages or spectra.
Note that the GW strains are always gauge invariant, $h_{ij}^{\rm TT} (\xx, t)$,
while the unprojected strains $h_{ij} (\xx, t)$ are a mathematical artifact used
to solve the GW equation.
Thus, we solve the linearised GW equation,
\EQ
\left( {\upartial^2 \over\upartial t^2}
- c^2\nabla^2 \right) h_{ij} (\xx, t)\, =\,
{\cal G}(t) \, T_{ij} (\xx, t)\,,
\label{d2hdt}
\EN
for the six components $1\leq i\leq j\leq3$.
Here, ${\cal G}(t)=6/t$ is the prefactor in the
normalised GW equation, given in (\ref{normalised_GW}),
that is only valid in the radiation-dominated
epoch, where the scale factor increases linearly with conformal time.
In the matter-dominated epoch, by contrast, there would be an additional
damping term that is not relevant for our study.
For test purposes, (\ref{d2hdt}) can also be applied to a non-expanding
universe by setting a constant ${\cal G}=6$; see
(\ref{GW_non_expanding}).
In this case, the scaled strains, comoving variables, and
conformal time, are readily given as physical quantities.

In the first approach, we solve (\ref{d2hdt}) using the third-order
Runge-Kutta scheme of \cite{Wil80}.
Furthermore, since the GW equation is second order in time, we also need
to advance the first conformal time derivative, $h'_{ij}(\xx, t)$.
Thus, the solution at the time $t + \delta t$, where $\delta t$ is the length
of the timestep, can be written in terms of the solution at the previous time
$t$ as
\begin{subequations}
\label{approachI_eq}
\EQ
\begin{pmatrix} h_{ij}\\ h'_{ij} \end{pmatrix}_{t+\delta t}
\,\equiv\,\qq_{3}\hskip 20mm \mbox{(approach~I)}\,,
\EN
where
\EQ
\qq_n\,=\,\qq_{n-1}+\beta_n\ww_n\,,\hskip 20mm
\ww_n\,=\,\alpha_n\ww_{n-1}+\delta t\QQ_{n-1}\,,
\EN
\end{subequations}
for $n = 1,\, 2, \,3$,
with $\alpha_1=0$, $\alpha_2=-5/9$, $\alpha_3=-153/128$,
$\beta_1=1/3$, $\beta_2=15/16$, $\beta_3=8/15$, and
\EQ
\qq_{n - 1}\,\equiv\,\begin{pmatrix} h_{ij}\\ h'_{ij} \end{pmatrix}_{n - 1}\,, \hskip 15mm
\QQ_{n - 1}\,\equiv\,\begin{pmatrix} h'_{ij}\\ c^2\nabla^2h_{ij}+{\cal G}T_{ij} \end{pmatrix}_{n - 1},
\label{approachI}
\EN
with initial $\qq_0$ and $\QQ_0$ evaluated at $t$.
This scheme is accurate to third order, i.e.,
$(h_{ij},\,h'_{ij})=(h_{ij},\,h'_{ij})_{\rm exact}+
c_{\rm Wil}\delta t^3(h_{ij},\,h'_{ij})'''$, where
$c_{\rm Wil}$ is a small number of the order of $10^{-4}$
which is characteristic of the Williamson scheme;
see \cite{Bra03} for various numerical tests.

Even in the general case when both ${\cal G} (t)$ and $T_{ij} (\xx, t)$ change
with time, those changes are small between two consecutive timesteps,
due to the restriction on $\delta t$ from the CFL condition.
In Fourier space, assuming the right-hand side to be constant in time
allows us to solve (\ref{d2hdt}) {\em exactly}
from one timestep to the next.
In this second approach, we first compute $\tilde{T}_{ij}^{\rm TT} (\kk, t)$
in Fourier space and then determine the two independent components in the
linear polarisation basis, $\tilde{T}_{+, \times} (\kk, t)$.
Thus, we only need to evolve $\tilde{h}_{+, \times} (\kk, t)$.
Knowing therefore $\tilde{h}_{+, \times} (\kk, t)$ and
$\tilde{h}'_{+, \times} (\kk, t)$ at the conformal
time $t$, we can compute the exact solution assuming constant ${\cal G} (t)
\tilde{T}_{+, \times} (\kk, t)$ over the duration of one timestep $(t, t + \delta t)$ as
\EQ
\begin{pmatrix}
\omega\tilde{h}-\omega^{-1}{\cal G}\tilde{T}\\
\tilde{h}'
\end{pmatrix}^{t+\delta t}_{+,\times}
=\begin{pmatrix}
~~\cos\omega\delta t & ~~\sin\omega\delta t \\
-\sin\omega\delta t  & ~~\cos\omega\delta t
\end{pmatrix}
\begin{pmatrix}
\omega\tilde{h}-\omega^{-1}{\cal G}\tilde{T}\\
\tilde{h}'
\end{pmatrix}^{t}_{+, \times}
\quad\quad\mbox{(approach~II)}.
\label{d2hdt2}
\EN
Here, $\omega=k$ with $k=|\kk|$ is the normalised angular frequency at
normalised wavevector $\kk$, according to the
normalisation defined in \App{normalised_eqn}.
In the following, both approaches will be compared with each other.

Approach~I involves numerical integration of (\ref{d2hdt}). In \Sec{results}, this will be seen
to produce numerical inaccuracy unless $\delta t$ is much smaller than the one
imposed by the CFL condition. Therefore, since the second approach circumvents this
drawback, the timestep is only limited by the CFL condition, which allows us to use a
much larger $\delta t$ and, hence, to compute $\tilde{T}_{+, \times} (\kk, t)$ at every
timestep, leading to a direct computation of the linear
polarisation modes $\tilde{h}_{+, \times}
(\kk, t)$. However, in approach~I, this is not computationally viable and, hence, we
compute $\tilde{h}_{+, \times} (\kk, t)$ from $h_{ij} (\xx, t)$
only when we are interested in obtaining the physical observables derived from the scaled strains.

We use the {\sc Pencil Code}\footnote{\url{https://github.com/pencil-code}, DOI:10.5281/zenodo.2315093}
for the numerical treatment of
(\ref{approachI_eq}) and (\ref{d2hdt2}) together with (\ref{dlnrhodt})--(\ref{dAdt}).
In its default configuration, it uses a sixth-order accurate discretisation
in space and a third-order accurate time stepping scheme.
Both, approaches~I and II are implemented as special modules
{\tt SPECIAL=gravitational\_waves\_hij6} and
{\tt gravitational\_waves\_hTXk}, respectively.

\section{Results}
\label{results}

\subsection{GWs for a Beltrami field}

It is useful to have an analytic solution to compare the numerical
solutions against.
A simple example that has not previously been discussed in this
context is the case of GWs generated by a magnetic Beltrami field
in a non-expanding flat universe, governed by (\ref{d2hdt}) with ${\cal G}=6$,
in the absence of fluid motions.
In this case, the scale factor, $a(t)$, does not affect the GW equation.
Hence, the initial time can be chosen to be zero.
The one-dimensional Beltrami magnetic field is expressed as
\begin{equation}
\BB(x, t)\,=\,B_0 \Theta(t)
\begin{pmatrix}
0\cr\sin k_0x\cr\cos k_0x
\end{pmatrix},
\label{Belt}
\end{equation}
where $k_0$ and $B_0$ are the characteristic wavenumber
and amplitude of the Beltrami field, respectively,
and $\Theta(t)$ is the Heaviside step function, such that the
sourcing magnetic field is assumed to appear abruptly at the
starting time of generation $t_*=0$.
In the present work, we assume this time
to be at the electroweak phase transition.
The normalised magnetic energy density,
$\varOmega_{\rm M} = B_0^2\left/(2 \EErad^*)\right.$
is constant in time.

The Beltrami field can equally well be applied to the velocity field, i.e.,
$\uu(x,t) = u_0 \Theta(t) \left( 0, \sin k_0 x, \cos k_0 x \right)^{\rm T}$.
The normalised kinetic energy density is
$\varOmega_{\rm K} = \rho u_0^2\left/(2 \EErad^*)\right.$.
In this case, there would be no initial magnetic field, although it could
be generated by a dynamo at later times, when $\eta \neq 0$. Hence, this case would require
solving the time-dependent MHD equations simultaneously with the GW equation,
if $\eta = 0$ cannot be assumed.

The fractional helicity of the Beltrami field is $\pm 1$ and has the same
sign as the characteristic wavenumber $k_0$.
The Beltrami field (when applied to $\BB$) is force-free
($\JJ\times\BB=\bm{0}$), so no velocity will be generated.
In the absence of magnetic diffusion ($\eta=0$), we can therefore treat
this magnetic field as given and do not need to evolve it.
In the TT projection, we can write the normalised stress tensor as
$T_{ij}^{\rm TT}(x, t) = (- B_iB_j + \half\delta_{ij}\BB^2)/\EErad^\ast$
for $i,j=2,3$ and $T_{ij}^{\rm TT}=0$ for $i=1$ and/or $j=1$.
We have
\begin{equation}
T_{ij}^{\rm TT}(x, t)\, = \,-\,\varOmega_{\rm M} \Theta (t)
\begin{pmatrix}
\,0 \,&\,     0    \,& \,    0 \,  \\[0.1em]
\,0 \,&\, -\cos2k_0x \,&\, \sin2k_0x\, \\[0.1em]
\,0 \,&\,~~\sin2k_0x \,&\, \cos2k_0x\,
\end{pmatrix}.
\label{T_ij_belt}
\end{equation}
For kinetic motions, the normalised stress-energy tensor can be written as
$T_{ij}^{\rm TT}(x, t) = \left( 4 \gamma^2 u_i u_j/c^2 + \delta_{ij} \right)
\Theta(t) \rho \left/\left(3 \rho_\ast \right)\right.
\sim \left[ 4 u_i u_j/c^2 \left( 1 + \half u_0^2/c^2 \right) +
\delta_{ij} \right] \Theta(t) \rho \left/\left(3 \rho_\ast \right)\right.$,
for $i,j=2,3$ and $T_{ij}^{\rm TT}=0$ for $i=1$ and/or $j=1$.
The contributions to the stress-energy tensor due to viscosity have been neglected.
Constant ($k = 0$) terms do
not source GWs.
Now, since $\rho$ is constant (note that the velocity field
defined is divergence-free, i.e., $\pmb{\nabla} {\bm \cdot} \uu = 0$), the
$\delta_{ij}$ terms can be neglected since they would be ruled out after linearisation.
Similarly, we can simplify $\sin^2 k_0 x = \frac{1}{2} \left( 1 - \cos 2k_0 x \right)$,
of which only the term $-\frac{1}{2} \cos 2 k_0 x$ gives a contribution,
which is at $k=2\,k_0$.
Likewise, $\cos^2 k_0 x$ leads only to a contribution given by $\frac{1}{2} \sin 2k_0 x$.
For this same reason, we have explicitly specified $T_{ij}^{\rm TT} = 0$
for $i = 1$ and/or $j = 1$, in both magnetic and kinetic Beltrami fields,
since the Beltrami fields defined are only functions of $x$, and
$T_{ij}^{\rm TT}$ is required to be transverse.
After these modifications, the normalised transverse and
traceless stress-energy tensor for kinetic motions is
\begin{equation}
T_{ij}^{\rm TT}(x, t)\,=\,{\textstyle\frac{4}{3}}\varOmega_{\rm K}
\left(1 + \half u_0^2/c^2 \right)
\Theta(t)
\begin{pmatrix}
\,0\, & \,    0   \, &\,     0 \,  \\[0.1em]
\,0\, &\, -\cos2k_0x\, &\, \sin2k_0x \,\\[0.1em]
\,0\, &\,~~\sin2k_0x\, &\, \cos2k_0x\,
\end{pmatrix},
\label{T_ij_belt_k}
\end{equation}
which is equivalent to (\ref{T_ij_belt}) with a different prefactor and opposite
sign.
From now on, we only refer to magnetic fields,
because the initial kinetic Beltrami field can lead
to velocity evolution and to magnetic field generation,
unless $\nu = \eta = 0$.
This would require solving the full system of MHD equations
and does not allow us to obtain simple analytic expressions as
desired for validation of the {\sc Pencil Code}.
If $\nu = \eta = 0$, the one-dimensional kinetic Beltrami field is a
stationary solution, and the results obtained for the magnetic field can
be applied to the kinetic field by changing $\varOmega_{\rm M}
\rightarrow - \frac{4}{3} \varOmega_{\rm K} \left(1 + \half u_0^2/c^2
\right)$.

Note that $T_{ij}^{\rm TT} (x, t)$ in (\ref{T_ij_belt}) and (\ref{T_ij_belt_k})
has only two independent terms, so we can directly compute the $+$ and
$\times$ components. For the Beltrami magnetic field, we have
\EQ
T_+ (x, t)\,=\,\varOmega_{\rm M}\Theta(t) \cos2k_0x,\hskip 10mm
T_\times (x, t)\,=\,-\,\varOmega_{\rm M}\Theta(t) \sin2k_0x\,.
\EN
These modes are directly obtained using the decomposition into
the $+, \times$ polarisation basis, described in \Sec{TT}, with
the change of sign described in \Sec{SignSwap}. If the change
of sign is not taken into account, the $\times$ mode obtained
is $\tilde{T}_\times (x, t) \propto {\rm i} \cos 2k_0 x$, which is
not independent of the $+$ mode. Since the $+$ and
$\times$ modes have to be orthogonal functions, the change of
sign is required to appropriately describe the modes.

Assuming $h_+=h_\times=\dot{h}_+=\dot{h}_\times=0$ at the initial time
$t=0$, when the Beltrami field starts to act as a source of GWs,
the time-dependent part of the solutions to (\ref{GW_non_expanding})
is proportional to $1-\cos 2 \omega_0 t = 2\sin^2\! \omega_0 t$,
where $\omega_0 = c k_0$, so we have
\begin{subequations}
\begin{align}
h_+(x,t)\,=&\hskip 2mm \frac{3 H_\ast^2}{c^2k_0^2}\,
\varOmega_{\rm M} \blue{\varTheta (t)}\cos2k_0x\,\sin^2\!\omega_0 t\,, \\
h_\times(x,t)\,=\,-&\hskip 2mm \frac{3 H_\ast^2}{c^2k_0^2}\,
\varOmega_{\rm M}\blue{\varTheta (t)} \sin2k_0x\,\sin^2\!\omega_0 t\,.
\end{align}
\end{subequations}
The spectral function $S_h$, defined in (\ref{Sh_app}), is given by
\EQ
S_h (k, t) \,= \,\left(\frac{3 H_*^2}{c^2 k_0^2}\right)^2 \varOmega^2_{\rm M}
\blue{\varTheta (t)}
\delta (k - 2\,k_0) \sin^4 \omega_0 \blue{t} \,,
\label{Sh_Bel}
\EN
where $\delta(k - 2\,k_0)$ is the Dirac delta function,
and the shell-integration is performed in 1D, such that we get a factor
$\varOmega_1 = 2$ in the computation.
The characteristic wavenumber $k_0$ has been considered to be positive because
the shell-integration rules out the dependence on direction of the wavevector $\kk$,
leading to a function that only depends on the positive modulus $k$. For
negative $k_0$ the $\delta$ term in (\ref{Sh_Bel}) should then be $\delta(k + 2k_0)$
instead.
\blue{In general, we write $\delta(k - 2|k_0|)$.}
The $\hrms (k, t)$ spectral function is
\begin{equation}
\hrms (k, t)\, =\, \frac{3 \sqrt{2} H_*^2}{c^2 k_0^2}  \sqrt{k_0} \varOmega_{\rm M}
\blue{\varTheta (t)}
\delta(k - 2\,\blue{|k_0|}) \sin^2 \omega_0 t\,,
\end{equation}
which leads to a characteristic amplitude
\begin{equation}
\hrms (t) \,=\, \left( \int_0^{\infty} S_h (k, t) \, \dd k \right)^{\!1/2} \,=\,
\frac{3 H_*^2}{c^2 k_0^2} \varOmega_{\rm M} \blue{\varTheta (t)}
\sin^2 \omega_0 t\,.
\label{hrms}
\end{equation}
The spectral function $S_{\dot{h}} (k, t)$ of (\ref{Sdoth_app}) is
given by
\begin{equation}
S_{\dot{h}} (k, t) \,= \,\left(\frac{3H_*^2}{c k_0} \right)^{\!2} \varOmega_{\rm M}^2
\blue{\varTheta (t)}
\delta (k - 2\,\blue{|k_0|})  \sin^2 2 \omega_0 t\,,
\end{equation}
where we have used the time derivatives of the strains
\begin{subequations}
\begin{align}
\dot{h}_+ (x, t)\, =&\hskip 2mm \frac{3 H_*^2}{c k_0} \varOmega_{\rm M} \blue{\varTheta (t)} \cos 2k_0 x
\sin 2 \omega_0 t, \\
\dot{h}_\times (x, t)\, =& \hskip 2mm \frac{3 H_*^2}{c k_0}
\varOmega_{\rm M} \blue{\varTheta (t)} \sin 2k_0 x \sin 2 \omega_0 t\,.\hskip 8mm
\end{align}
\end{subequations}
The normalised GW energy spectrum $\varOmega_{\rm GW} (k, t)$ is given by
\EQ
\varOmega_{\rm GW} (k, t)\, =\, \frac{k S_{\dot{h}}}{6 H_*^2} \,=\,
\blue{\frac{3 H_\ast^2}{c^2 |k_0|}} \varOmega_{\rm M}^2 \blue{\varTheta (t)}
\delta (k - \blue{2|k_0|})
\sin^2 2\omega_0 t\,.
\label{EEGW_fourier}
\EN
This leads to the total normalised GW energy density
\begin{equation}
\varOmega_{\rm GW} (t) \,=\, \frac{1}{6H_*^2} \int_0^{\infty} S_{\dot{h}} (k, t)\, \dd k\,
= \,\blue{\frac{3}{2}}\frac{H_*^2}{c^2 k_0^2} \varOmega_{\rm M}^2
\blue{\varTheta (t)} \sin^2 2 \omega_0t\,.
\label{varOmega_GW}
\end{equation}
The time-averaged values of $\hrms (t)$
and $\varOmega_{\rm GW} (t)$ are given by
\EQ
\bar{h}_{\rm rms}\,=\,\blue{\frac{3}{2}} \frac{H_\ast^2}{c^2 k_0^2}\,
\varOmega_{\rm M}\,,\hskip 15mm
\bar{\varOmega}_{\rm GW}\, =\, \blue{\frac{3}{4}} \frac{H_\ast^2}{c^2 k_0^2}\,\varOmega_{\rm M}^2\,.
\label{hrmsBF_EEGWBF}
\EN
Note that the energy ratio obeys $\bar{\varOmega}_{\rm GW}/
\varOmega_{\rm M}=\, \blue{\bar{h}_{\rm rms}/2}$
and is thus proportional to $k_0^{-2}$.
The normalised antisymmetric spectral function
$A_{\dot{h}} (k, t)$ of (\ref{Adoth_app}) is
given by
\begin{equation}
A_{\dot{h}} (k, t)\, =\, \left( \frac{3 H_*^2}{ck_0} \right)^2
\varOmega_{\rm M}^2 \blue{\varTheta (t)} \delta (k - 2k_0) \sin^2 2 \omega_0 t\,,
\label{Ah_Bel}
\end{equation}
where $k_0$ has been considered to be positive. For negative $k_0$
the $\delta$ factor in (\ref{Ah_Bel}) changes to $- \delta(k + 2\,k_0)$,
where the negative sign corresponds to negative $\pmb{k}$,
as described in \Sec{SignSwap}.
\blue{In general, we write $\sgn(k_0) \delta(k - 2|k_0|)$}.
The normalised antisymmetric spectral GW function
$\varXi_{\rm GW}(k,t)$ is given in Fourier space by
\EQ
\varXi_{\rm GW}(k, t)\, =\, \frac{k A_{\dot{h}}}{6 H_*^2}\, =\,
\blue{\frac{3H_\ast^2}{c^2 k_0}} \varOmega_{\rm M}^2 \blue{\varTheta (t)}
\blue{ \sgn(k_0) \delta (k - 2|k_0|)} \sin^2 2 \omega_0 t\,.
\label{A_GW}
\EN
This leads to the total normalised $\varXi_{\rm GW} (t)$
\begin{equation}
\varXi_{\rm GW} (t) \,=\, \frac{1}{6H_*^2} \int_0^{\infty} A_{\dot{h}} (k, t) \dd k \, =\,
\blue{\frac{3}{2}}\frac{H_*^2}{c^2 k_0^2} \varOmega_{\rm M}^2
\blue{\varTheta (t) \sgn(k_0)} \sin^2 2 \omega_0 t\,,
\end{equation}
which corresponds to the same function of time as $\varOmega_{\rm GW} (t)$.
The degree of circular polarisation is obtained as the fraction of
the antisymmetric spectral function to the GW energy density,
$\mathcal{P}_{\rm GW}(k, t) = \varXi_{\rm GW} (k, t)\left/
\varOmega_{\rm GW} (k, t),\right.$
such that the polarisation degree of Beltrami fields is $\pm 1$
at $k = \pm 2k_0$ and the sign is the same
as the sign of the magnetic helicity, given by the sign of $k_0$,
and undefined for other values of $k$.
The characteristic wavenumber of GWs is $k_{\rm GW} = 2k_0$ for the Beltrami field,
where $k_0$ is the characteristic wavenumber of the source $E_{\rm M} (k, t)$.
Hence, we observe a shift between the characteristic wavenumber of
$\varOmega_{\rm GW} (k, t)$ and that of $\varOmega_{\rm M} (k, t)$.

\subsection{Numerical solutions for approach~I
at finite spatio-temporal resolution}
\label{FiniteResolution}

At small enough grid spacings and small enough timesteps, our numerical
solutions reproduce the considered
one-dimensional Beltrami field.
At coarser resolution, however, we find that $\hrms(t)$ and
$\varOmega_{\rm GW} (t)$ are characterised by an additional decay of the form
\begin{subequations}
\begin{align}
\hrms(t)\,=\,&\,\Big( 3 H_\ast^2\left/ c^2 k_0^2 \right. \Big) \varOmega_{\rm M}
\, {\mathrm e}^{-\lambda t} \sin^2 \omega_0 t\,, \\
\varOmega_{\rm GW} (t) \,=\,& \,\blue{\Big(3 H_\ast^2\left/ 2 c^2 k_0^2 \right. \Big)}
\varOmega_{\rm M}^2
\, {\mathrm e}^{-\lambda t} \sin^2 2 \omega_0 t\,,
\end{align}
\end{subequations}
where $\lambda$ is the numerical decay rate.
We emphasise that $\lambda\neq0$ is entirely artificial and has to do
with imperfect numerics in the case of approach~I.
Results for $\lambda$ are given in \Tab{Tres} as functions of $\delta t$
(quantified by the Courant number $\delta t\,c/\delta x$) and the GW wavenumber
$\kGW$ (normalised by the Nyquist wavenumber $\kNy=\pi/\delta x$ to give
$\kGW/\kNy=\kGW\delta x/\pi$), defined in (\ref{kGW1}).

\begin{figure*}[t!]\begin{center}
\includegraphics[width=.7\textwidth]{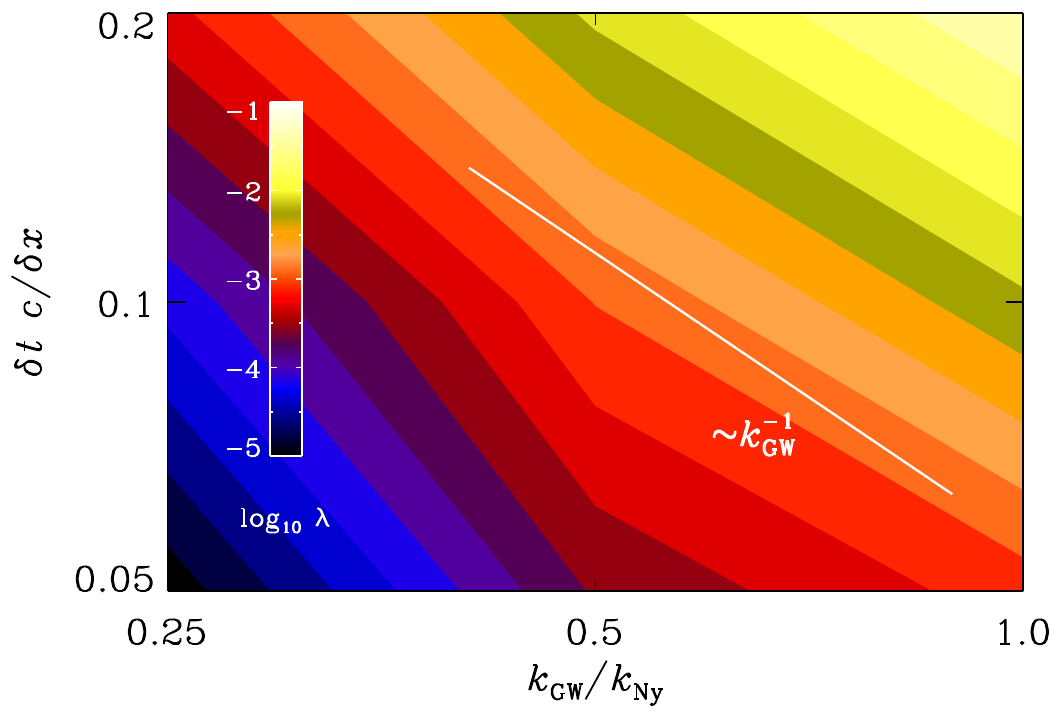}
\end{center}\caption[]{
Dependence of the decay rate of the Beltrami field solution due to
numerical error, $\lambda$, on $\kGW/\kNy$ and $\delta t\,c/\delta x$.
Blue (yellow) shades indicate low (high) numerical errors.
The error is high when $\kGW$ is close to the Nyquist wavenumber
and $\delta t$ is large.
The scaling $\delta t\,c/\delta x \propto (\kGW/\kNy)^{-1}$ is
indicated by a white solid line (colour online).
}\label{pres}\end{figure*}

\begin{figure*}[t!]\begin{center}
\includegraphics[width=\textwidth]{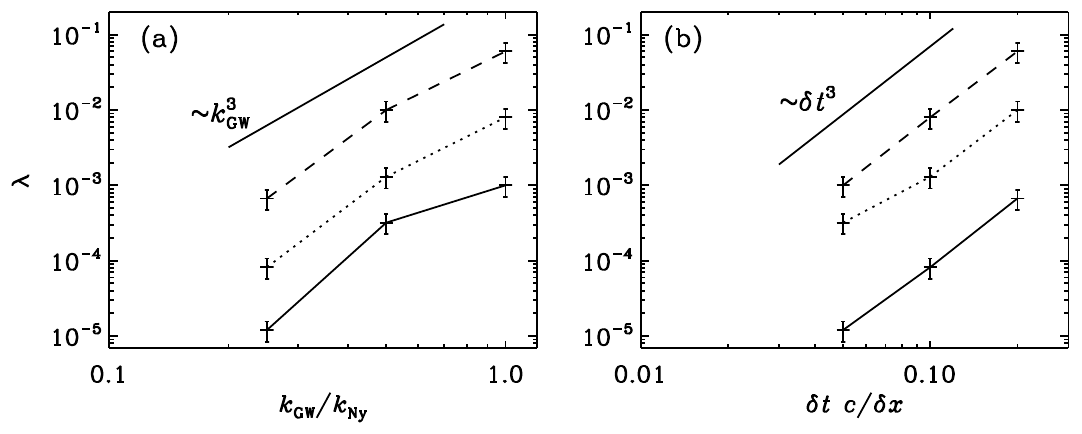}
\end{center}\caption[]{
Scaling of the decay rate of the Beltrami field solution due to
numerical error, $\lambda$, with $\kGW/\kNy$ (left panel) for
$\delta t\,c/\delta x = 0.05$ (straight line),
$0.1$ (dotted line), and $0.2$ (dashed line), and with
$\delta t\,c/\delta x$ (right panel) for
$\kGW/\kNy = 0.25$ (straight line), $0.5$ (dotted line), and $1.0$
(dashed line).
Cubic scalings are indicated by the straight lines.
}\label{pscl}\end{figure*}

The magnetic field wavenumber, see (\ref{kM}), is $k_{\rm M} = k_0$,
so $\kGW=2k_0 \ = 2k_{\rm M}$ is the wavenumber of the GWs generated by the
one-dimensional Beltrami field.
We see from \Tab{Tres} that the decay rate is largest
for $\kGW=\kNy$ and varies there between $6\times10^{-2}$
(for $\delta t\,c/\delta x=0.2$)
and $10^{-3}$ (for $\delta t\,c/\delta x=0.05$).
In \FFig{pres} we plot contours of $\lambda$ (colour-coded) versus $\kGW$
and $\delta t$.
Again, the largest values of $\lambda$ occur when $\kGW=\kNy$ and $\delta t$ is
large.
We also see that the lines of constant decay rate scale like
$\delta t\propto\kGW^{-1}$.
In \Fig{pscl} we show, in separate panels, the changes in $\lambda$ versus $\kGW$ and $\delta t$.
We see that the data points are compatible with the scalings
$\lambda\propto\kGW^3$ and $\lambda\propto\delta t^3$.
The cubic scaling of $\lambda$ is related to the
third order accuracy of the time stepping scheme.
The slight departures from this behaviour can be attributed
to the low number of runs computed to construct \Tab{Tres}.

\begin{table}[t!]\caption{
Dependence of the numerical decay rate $\lambda$
on $\kGW/\kNy$ and $\delta t\,c/\delta x$.
}\vspace{12pt}\centerline{\begin{tabular}{lccccccc}
$\delta t\,c/\delta x$ &  $\kGW/\kNy=1/4$  &  1/2  &  1  \\
\hline
0.2  &  $6.7\times10^{-4}$  &  $1.0\times10^{-2}$  & $6.0\times10^{-2}$  \\
0.1  &  $8.2\times10^{-5}$  &  $1.3\times10^{-3}$  & $8.0\times10^{-3}$  \\
0.05 &  $1.2\times10^{-5}$  &  $3.2\times10^{-4}$  & $1.0\times10^{-3}$  \\
\label{Tres}\end{tabular}}\end{table}

\subsection{Timestep constraint for approach~I in a turbulent case}

We now present an example where the timestep constraint becomes
particularly apparent when directly integrating the GW equation.
As alluded to in the introduction, this is the case when GWs are being
sourced by turbulent stresses, and we use the approach~I.
We consider here the case of decaying helical magnetic turbulence.
This case was originally considered in the cosmological context
using just an irregular magnetic field and no flow as the initial
condition \citep{BEO96,CHB01,BJ04}.
In this context, one can argue that the magnetic field at scales
larger than the injection scale must be causally related.
This, together with the solenoidality of the magnetic field,
leads to a $k^4$ subinertial range spectrum \citep{DC03}.
The $E_{\rm M}(k, t)\propto k^4$ scaling corresponds to
$\varOmega_{\rm M} (k, t)\propto k^5$.
The magnetic field is strong and the fluid motions are just the
result of the Lorentz force.
The $k^4$ subinertial range spectrum is then followed by a
$k^{-2}$ weak turbulence spectrum at high wavenumbers \citep{BKT15}.
The normalised wavenumber where the change of behaviour occurs, is the peak
wavenumber, $k_*$.
For an initial $k^4$ spectrum, the magnetic field undergoes inverse
cascading such that the magnetic energy spectrum is self-similar and obeys
\EQ
\varOmega_{\rm M}(k, t)/k\,=\,\phiM\left(k\xiM(t)\right),
\label{phiM}
\EN
where $\phiM$ is a generic function \citep{CHB01,BK17} and
$\xiM(t)$ is the magnetic integral scale given by (\ref{kM}).

\begin{figure*}[t!]\begin{center}
\includegraphics[width=\textwidth]{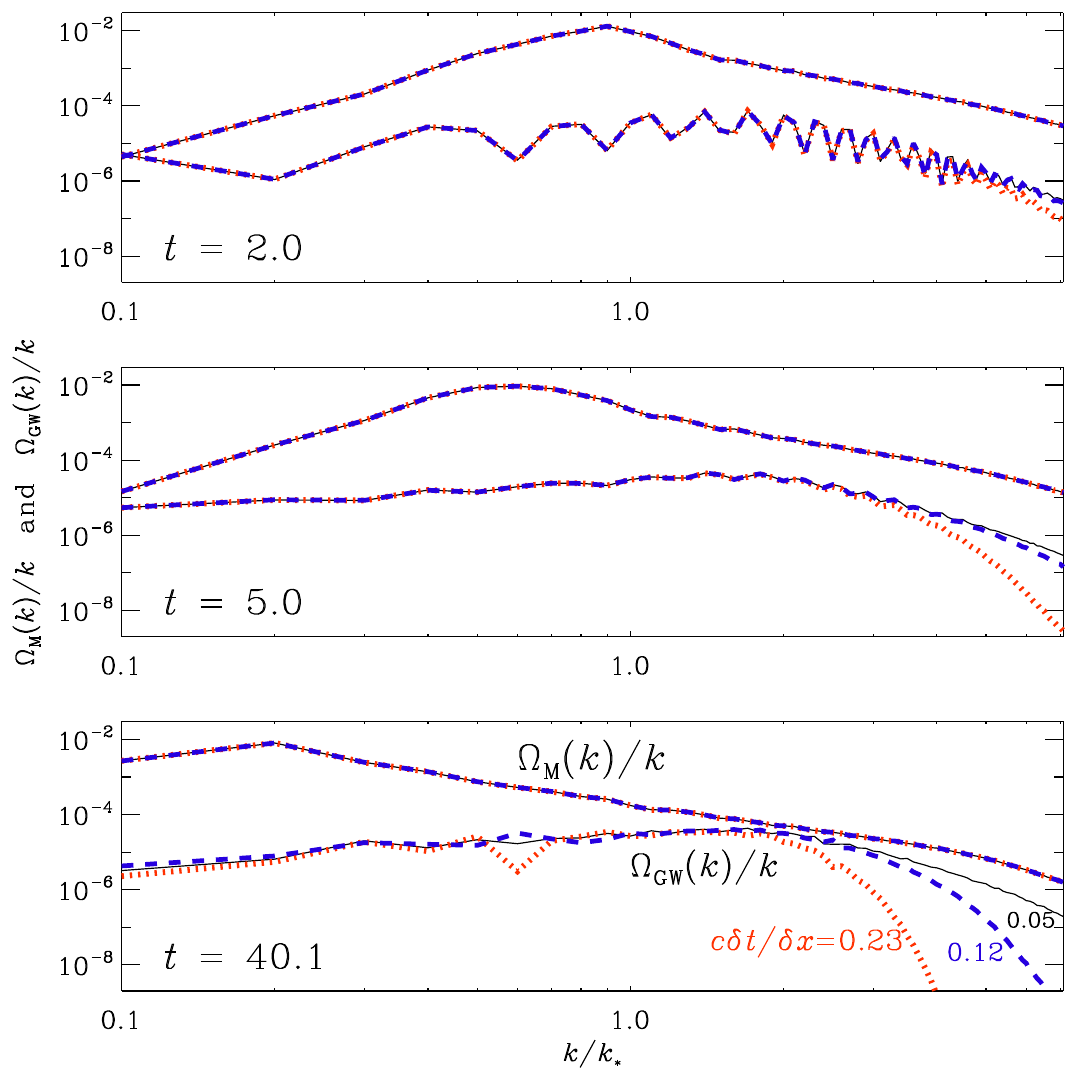}
\end{center}\caption[]{
GW (lower curves) and magnetic (upper curves) energy
spectra at three different normalised conformal
times, $t=2$ (top panel), $5$ (middle
panel), and $40$ (bottom panel), which are normalised with the
time when the turbulence source turns on, i.e., $t_*$.
Each panel shows three different
Courant numbers:  $\delta t\,c/\delta x=0.23$ (red dotted lines),  0.12 (blue
dashed lines), and 0.05 (black solid lines).
The wavenumbers are normalised with the normalised peak wavenumber $k_\ast$, at the
starting time of
generation.  In this simulation, $k_\ast = 10$, and
the total initial magnetic field energy density is $\varOmega_{\rm M} \, (t = 1) \approx
0.123$ (colour online).
}\label{ppower_comp}\end{figure*}

In \Fig{ppower_comp} we show, for three different times,
the normalised magnetic and GW energy
spectra, obtained following approach~I, for an expanding universe,
so we have ${\cal G} (t) =6/t$
on the right-hand side of (\ref{d2hdt}), and $t=1$ is the initial
normalised time, which refers to the starting time of generation.
Independent of the value of $\delta t$, the peak of
$\varOmega_{\rm M}(k,t)/k$ is seen
to propagate gradually towards smaller $k$.
This is the inverse cascade owing to the presence of magnetic helicity
\citep{PFL76,BM99,CHB01}.
Note that the peak of the spectrum always has the same height.
This is compatible with (\ref{phiM}).
The ratio of $\varOmega_{\rm M} (t) /\varOmega_{\rm GW} (t)$
changes from about 100 at early times to about 20 at the last
time as the magnetic field decays, while $\varOmega_{\rm GW} (t)$
stays approximately constant.

Let us now focus on the comparison of solutions for different
Courant numbers,
$\delta t\,c/\delta x=0.23$, $0.12$, and $0.05$ in \Fig{ppower_comp}.
While the magnetic energy spectra are virtually identical for different
$\delta t$, even for high wavenumbers, the GW spectra are not.
There is a dramatic loss of power at large $k$, when $\delta t\,c/\delta x=0.23$.
A value of $\delta t\,c/\delta x=0.8$ was always found to be safe as
far as the hydrodynamics is concerned, but this is obviously not small enough
for the GW solution.
This is a surprising result that may not have been noted previously.

\begin{figure*}[t!]\begin{center}
\includegraphics[width=\textwidth]{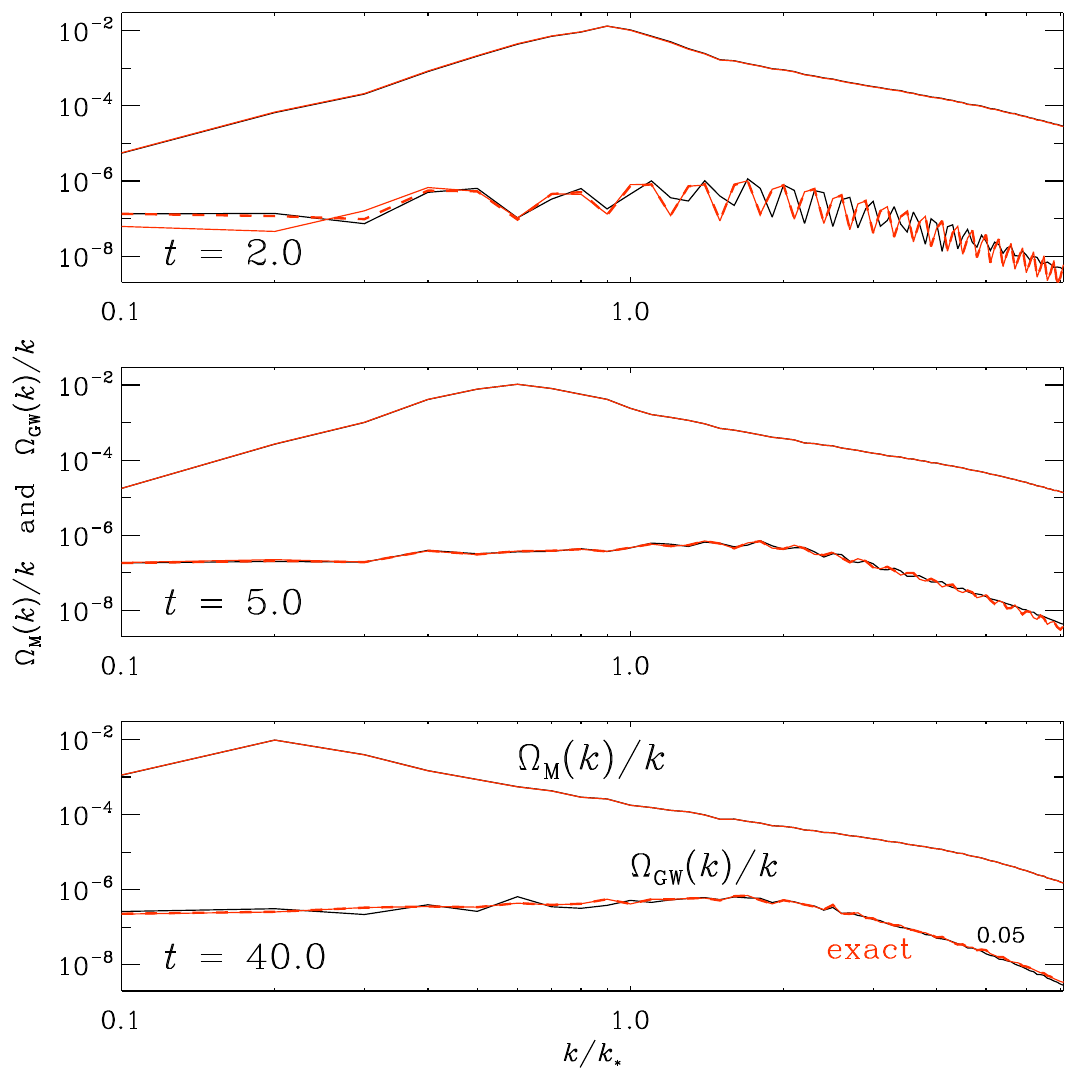}
\end{center}\caption[]{
Same as \Fig{ppower_comp}, but now comparing the case with
$\delta t\,c/\delta x=0.05$ (black solid lines), using approach~I, with the analytic solution
assuming constant ${\cal G}(t) T_{ij} (\xx, t)$, in time,
between consecutive timesteps
(red solid line), which corresponds to approach ~II, here referred to
as ``exact'', using $\delta t\,c/\delta x=0.8$.
The red dashed lines indicate the spectrum computed from the
conformal time derivatives of the strains, $\varOmega^{\rm h'}_{\rm GW}(k,t)/k$  (colour online).
}\label{ppower_comp6k}\end{figure*}

\begin{table}[t!]\caption{
Dependence of the decay rate of the numerical error $\lambda$
on $\kGW/\kNy$ and $\delta t\,c/\delta x$
for hydromagnetically driven GWs.
Dashes indicate that the decay rate was too small compared with
the fluctuations and could not be determined.
}\vspace{12pt}\centerline{\begin{tabular}{lccccccc}
$\delta t\,c/\delta x$ &  $\kGW/\kNy=1/4$  &  1/2  &  1  \\
\hline
0.23 &  0.012  &  0.16  & 1.1  \\
0.12 &    --   &  0.015 & 0.12 \\
0.05 &    --   &    --  & 0.006 \\
\label{TresDNS}\end{tabular}}\end{table}

In \Fig{ppower_comp6k}, we show that the solution with
$\delta t\,c/\delta x\la0.05$ agrees perfectly with that of
approach~II at $\delta t\,c/\delta x\la0.8$.
The additional cost in performing Fourier transforms at every
timestep, used in approach~II, is easily outweighted by the more than 10 times
longer timestep, when comparing to approach~I.
We also show the comparison between the actual GW energy
spectrum, $\varOmega_{\rm GW} (k, t)$; see (\ref{OmegaGW_app}), and the spectrum obtained
from conformal time derivatives, $\varOmega^{h'}_{\rm GW}(k, t)$;
see (\ref{Omega_GW_h'_app}).
These two spectra become more similar for large wavenumbers and
for longer times.

To see whether the observed degradation using approach~I; see \Fig{ppower_comp},
is compatible with what has
been seen for the monochromatic Beltrami field, we determine again
the decay rates for three wavenumbers of the spectral GW amplitude.
The result is given in \Tab{TresDNS}.
We see that, although the scalings with $\kGW$ and $\delta t$ are
compatible with what has been seen in \Sec{FiniteResolution} for the
Beltrami field, the actual values of $\lambda$ are about 12 times larger.
The reason for this is not clear at this point.

\begin{figure*}[t!]\begin{center}
\includegraphics[width=\textwidth]{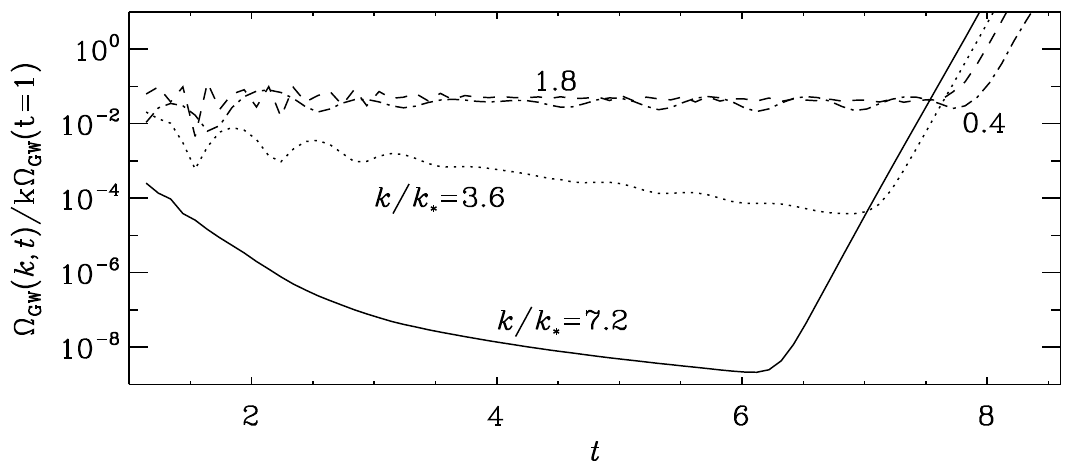}
\end{center}\caption[]{
Late time numerical instability for $\delta t\,c/\delta x=0.46$ seen in
the temporal evolution of $\varOmega_{\rm GW}(k,t)/k$
after $t=6$ for $k/k_*=7.2$
(solid), 3.6 (dotted), 1.8 (dashed), and 0.4 (dash-dotted).
The spectra shown are normalised with the total GW energy density shortly
after the start of the simulation, at $t \ga 1$,
$\varOmega_{\rm GW} (k, t)/k \varOmega_{\rm GW} (t \ga 1)$.
The normalised peak wavenumber is $k_\ast = 10$, and the total initial
magnetic energy density is $\varOmega_{\rm M} \, (t = 1) \approx
0.123$.
}\label{plate_growth}\end{figure*}

In addition to the numerical error discussed above, when computing the
solution using approach~I, we found
a numerical instability that is distinct from the usual
one invoked in connection with the CFL condition.
This new instability emerges when the accuracy of
the solution is already strongly affected by the length of the timestep,
namely for $\delta t\,c/\delta x\, \ga\, 0.46$, which is still well within
the range of what would normally (in hydrodynamics) be numerically stable.
The problem appears at late times, after the GW spectrum has long
been established.
This new numerical error manifests itself as an exponential growth that
is seen first at large wavenumbers and then at progressively smaller ones;
see \Fig{plate_growth}.
Our earlier studies have shown that this problem cannot be controlled by adding
explicit diffusion to the GW equation.
Given that the solution is already no longer accurate for this length
of the timestep, this numerical instability was not worth further
investigation, but it highlights once again the surprising differences
in the numerical behaviour of wave and fluid equations.

\begin{figure*}[t!]\begin{center}
\includegraphics[width=\textwidth]{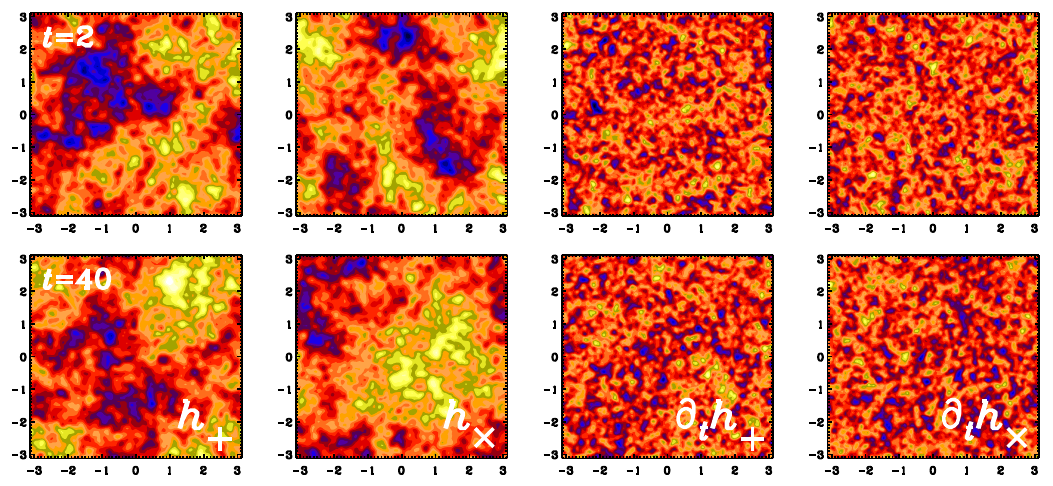}
\end{center}\caption[]{
$xy$ cross-sections through $z=0$ of $\hT (\xx, t)$,
$\hX (\xx, t)$, and the normalised conformal time derivatives,
$\hT'(\xx, t) = \upartial_t \hT (\xx, t)$, and $\hX' (\xx, t)
= \upartial_t \hT (\xx, t)$,
at $t=2$ (upper row) and $t=40$ (lower row).
The simulation parameters are the same as in \Figs{ppower_comp}{ppower_comp6k} (colour online).
}\label{pslices}\end{figure*}

Looking again at the GW energy spectra at early times,
\Figs{ppower_comp}{ppower_comp6k}, we see
wiggles in the spectrum at $t=2$ (top panel).
One might be concerned that these are caused by numerical artifacts,
but the spatial distributions of physical $\hT (\xx, t = 2)$ and
$\hX (\xx, t = 2)$ look smooth; see \Fig{pslices}.
Thus, the wiggles in $\varOmega_{\rm GW}(k,t)/k$
are not artificial, but presumably related to
the finite domain size and the way the initial condition
for the magnetic field, $\varOmega_{\rm M} \, (k, t = 1)/k$
is posed using combined $k^4$ and $k^{-2}$
power laws, for small and large wavenumbers respectively.
They might appear as a transient effect in the evolution from the initially
vanishing GW energy density to the shape observed for later times.
Indeed, at late times the wiggles disappear.

\section{Can the timestep cause artifacts in hydrodynamic and MHD turbulence?}

There have been reports in the literature that the length of
the timestep can affect the convergence properties of solutions of
incompressible hydrodynamic simulations \citep{SKKRD15}.
In the present compressible MHD simulations, however, no obvious side
effects of increasing the length of the timestep within the standard
CFL condition have been seen.
However, there could be subtle effects.
Here we investigate two possibilities.
The first is the bottleneck effect in hydrodynamic turbulence,
which refers to the kinetic energy spectrum,
$\varOmega_{\rm K}(k, t)/k$, slightly shallower than the Kolmogorov
$k^{-5/3}$ spectrum.
This phenomenon is explained by the inability of triad interactions
with modes in the dissipative subrange to dispose of turbulent energy
from the end of the inertial range \citep{Fal94}.
This also has subtle effects on the growth rate of turbulent small-scale
dynamos \citep[see][hereafter referred to as BHLS]{BHLS18}.
The second possible subtlety is a modification of the magnetic energy
spectrum, $\varOmega_{\rm M} (k, t)/k$, during the kinematic growth phase.
This problem of a kinematic small-scale dynamo is closest to our GW
experiment in that both problems are linear and there is no turbulent
cascade in either of the two problems.
We begin with the first possibility.

The description in this section refers only to MHD turbulence
and, for convenience, the usual non-normalised
and physical variables are used for
comparison with other works (e.g., $k$ refers to dimensional physical
wavenumbers), instead of the normalised variables that are
useful in the context of GWs.
As in previous sections, however, we continue to show magnetic and
kinetic energy spectra in terms of
$\varOmega_{\rm M, K} (k, t)/k$.

In the simulations of BHLS, turbulence was being forced at low wavenumbers
using an explicit forcing function $\ff(\xx,t)$ on
the momentum equation.
It drives modes in a narrow band of wavenumbers.
We consider here run~D of BHLS, where driving was applied at wavenumbers
between 1.4 and 1.8 times the lowest wavenumber of the domain,
$k_{\min}=2\pi/L$ of a cubic domain of size $L^3$.
The magnetic Reynolds number based on the average wavenumber
was about $540$.

\begin{figure*}[t!]\begin{center}
\includegraphics[width=\textwidth]{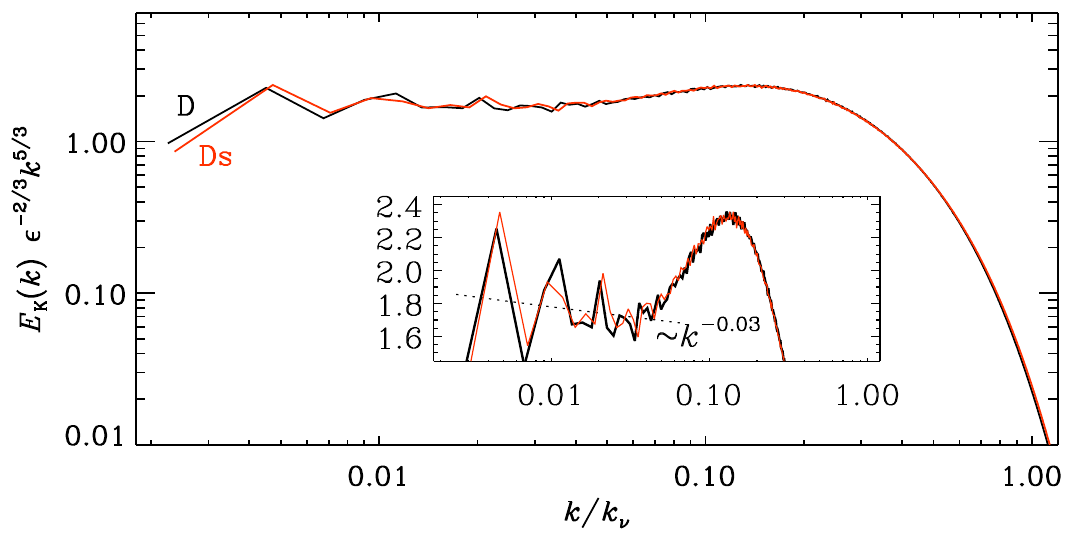}
\end{center}\caption[]{
Comparison of the kinetic spectrum compensated
with the usual Kolmogorov inertial scaling
$k^{-5/3} \epsilon_{\rm K}^{2/3}$, for run~D (black line)
of BHLS with those of a continuation of this run
with a four times shorter timestep (run~Ds, red line).
The inset shows the compensated spectra on a linear scale.
The dotted line shows the expected inertial range
correction proportional to $k^{-0.03}$ \citep{Anselmet,SL94}.
The physical wavenumbers are normalised with the dissipation wavenumber
$k_\nu$.
Both spectra are time-averaged after compensating against the
exponential growth (colour online).
}\label{pspecm_comp_short}\end{figure*}

The important point of BHLS was to show that the bottleneck effect is
independent of the forcing wavenumber, provided that the effective forcing
wavenumber is used in the definition of the magnetic Reynolds number.
Here we demonstrate that the bottleneck is not affected by the length
of the timestep.
Technically, the simulations presented in this section are done with magnetic
fields included, but the field
is at all scales still extremely weak, so for all practical purposes
we can consider those as hydrodynamic simulations.
The result is shown in \Fig{pspecm_comp_short}, where we compare run~D
of BHLS, which uses a timestep of $\delta t\,\cs/\delta x=0.6$,
with a new one called run~Ds, where `s' indicates that
the timestep is shorter, such that now $\delta t\,\cs/\delta x=0.15$,
where $\cs$ refers to the sound speed.
Both spectra fall off in the same way as $k$ approaches the
viscous cutoff wavenumber $k_\nu=(\epsilon_{\rm K}/\nu^3)^{1/4}$,
where $\epsilon_{\rm K}$ is the mean kinetic energy injection rate
per unit mass, and $\nu$ is the kinematic viscosity.

It turns out that there is no difference in the energy spectrum relative
to run D, where the timestep obeys $\delta t\,\cs/\delta x=0.60$.
Thus, the artifacts reported in the present paper, namely the excessive
damping of power at high wavenumbers, seem to be confined to the GW
spectrum and do not affect in any obvious way the properties of the
energy spectrum of MHD turbulence.

\begin{figure}\begin{center}
\includegraphics[width=\columnwidth]{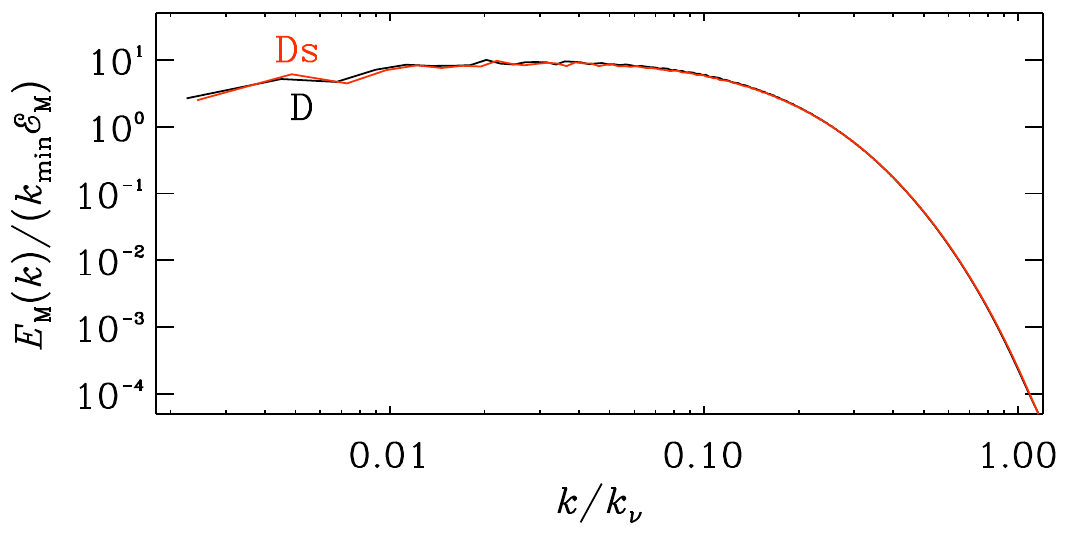}
\end{center}\caption[]{
Comparison of the magnetic energy spectrum, $\varOmega_{\rm M} (k)/k$,
normalised with the total magnetic energy $\EEM$, for run~D (black line)
of BHLS with that of a continuation of this run
with a four times shorter timestep (run~Ds, red line) (colour online).
}\label{pspecm_kazan_short}\end{figure}

\begin{figure}\begin{center}
\includegraphics[width=\columnwidth]{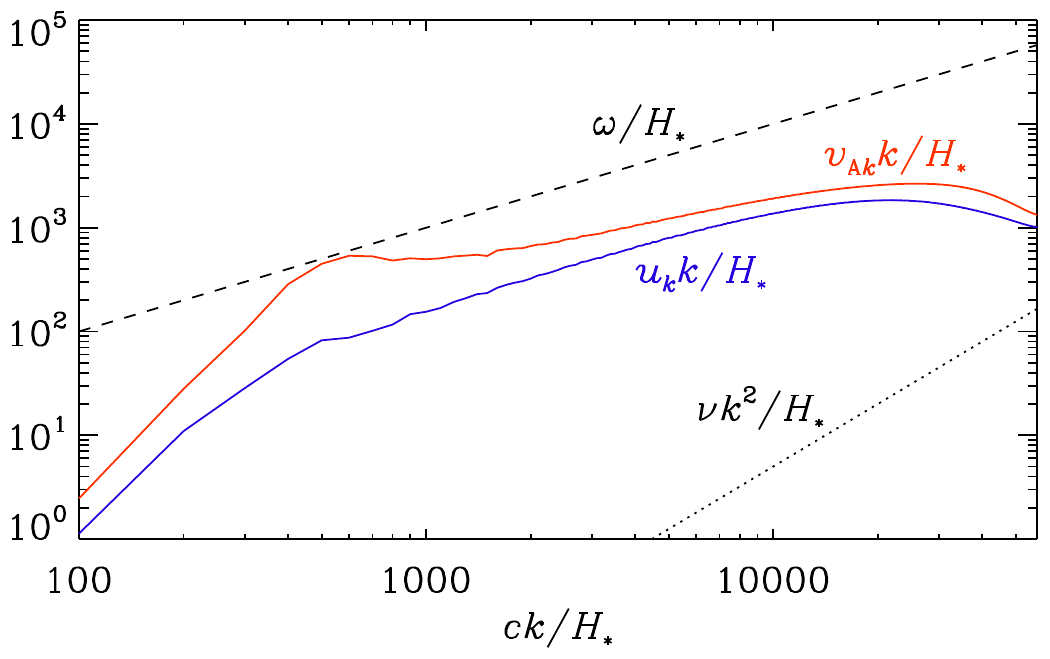}
\end{center}\caption[]{
Comparison of the GW frequency $\omega=ck$ with the
turnover rate $u_k (k)\, k$ of the turbulence, the turbulent
Alfv\'en rate $v_{{\rm A}k} (k)\, k$, and the viscous
damping rate $\nu k^2$ at normalised wavenumber $ck/H_*$.
Note that all frequencies in the plot are normalised with $H_*$ (colour online).
}\label{pomegas}\end{figure}

Next, let us look at the magnetic energy spectrum during the
kinematic stage.
Again, we compare run~D of BHLS with our new run~Ds.
The result is shown in \Fig{pspecm_kazan_short}, where we
see virtually no difference between the two curves.
In this sense, the idea that the sharp decay of the
MHD energy spectra at higher wavenumbers
is being masked by the forward cascade of energy may not be
borne out by the simulations.
However, it is possible that the spreading of energy
across wavenumbers is not so much the result of
nonlinearity, which is indeed unimportant in the kinematic
stage of a dynamo, but that it is due to the fact that
the induction equation has non-constant coefficients.
This leads to mode coupling, as has been seen in other
turbulent systems during the linear stage.
An example is the Bell instability \citep{Bell04},
where significant spreading of energy across different
modes has been observed \citep[see, for example, figure~4 of][]{RKBE12}.

The lack of any noticeable high wavenumber artifacts in MHD turbulence
can simply be explained by the absence of relatively rapid oscillations
in MHD flows, compared to GW oscillations, which are proportional to $c$.
To demonstrate this, we compare in \Fig{pomegas} the GW frequency
$\omega=ck$ with the turbulent turnover rate $u_k (k) k$,
the turbulent Alfv\'en rate $v_{{\rm A}k} (k) k$, and the viscous damping
rate $\nu k^2$ at wavenumber $k$ for a simulation of GWs at $1152^3$
mesh points, a Reynolds number, Re = $\urms/\nu k$ of about 1000, and a
magnetic Reynolds number Rm = Re, so that $\nu = \eta$.
Here, we use the relations $u_k (k) =\sqrt{2\EK(k)}$, and
$v_{{\rm A}k} (k) =\sqrt{2\EM(k)}$ for the $k$-dependent turbulent
velocity and the $k$-dependent Alfv\'en speed.
Note that $\max(v_{{\rm A}k} (k) \, k)$, at $c k/H_* \approx 3 \cdot 10^{-4}$,
is about 30 times smaller than $\omega = ck$, and
the difference is bigger for larger values of $k$.
This shows that from an accuracy point of view, the timestep could
well be 30 times longer before the accuracy of MHD begins to be affected.

\section{Conclusions}

Our work has identified an important aspect that constrains the length
of the timestep in numerical solutions of the linearised GW
equation through approach~I in a more stringent way than what is
known in the more familiar MHD context.
In this approach, the wave equation is integrated in the same way as
the MHD equations.
The resulting numerical error affects the GW spectrum at high wavenumbers in
a systematic way, which may have gone unnoticed in earlier investigations.

Similar numerical errors do not seem to affect the
hydromagnetic energy spectra,
which could be a consequence of a direct energy cascade.
It therefore appears that the direct energy cascade is insensitive to
numerical errors in hydrodynamic and hydromagnetic turbulence.
As discussed, fast turbulent MHD waves might have very small amplitudes
due to the viscous cutoff, which becomes dominant in the range of high
frequencies (see \Fig{pomegas}). On the other hand, this cutoff is not present in the GW
oscillations. This allows the presence of waves with very high
frequencies that require very small $\delta t$ to accurately
represent the high frequency oscillations numerically.
A particularly interesting aspect of numerical solutions of turbulence
is the bottleneck effect, i.e., a shallower fall-off of spectral energy
just before turbulence disposes of its energy in the viscous dissipative
subrange \citep[][BHLS]{Kaneda,HB06}.
One might therefore be concerned that this bottleneck itself could be
the result of numerical artifacts by a finite timestep.
However, we have now shown that this is not the case and that the
height of the bottleneck is independent of the length of the timestep.
Nevertheless, there could be other effects, especially in strongly
compressible turbulence, where short timescales play a role and the solution
might contain errors for timesteps below the CFL condition.

We conclude that, to obtain the correct energy spectrum at high
wavenumbers with approach~I, it is essential to compute solutions with
rather small timesteps of just a few percent of the CFL condition.
This makes reasonably accurate GW spectra (where the decay rate
is less than 1\% of the Hubble rate, i.e., the case with
$\delta t\,c/\delta x=0.05$ in \Fig{ppower_comp}) by at least
a factor of 10 more costly than what would naively be expected.
In multiphysics systems, as in our case, where we solve coupled
MHD equations and linearised Einstein field equations, it
becomes even more important to lower the upper bound on $\delta t$,
since this is affecting the duration of the numerical simulation
of the whole system of equations.
With approach~II, however, accurate solutions have been found at
ordinary timestep lengths with $C_{\rm CFL}=0.8$.
The less stringent requirement on accuracy is due to the fact that the
GW solution is not numerically integrated, and the oscillations are
represented analytically between subsequent timesteps. This allows the
representation of very high frequency oscillations, even if $\delta t$
is not small enough to numerically represent these oscillations accurately.
High resolution runs have now been carried out with this scheme for
a range of cases with initial magnetic field and also for cases where
hydrodynamic or hydromagnetic turbulence is driven during a brief episode
of stochastic forcing either in the momentum equation or in the induction
equation.
Those results are being reported in a separate publication \citep{RPMBKK19}.

\section*{Acknowledgements}
This work was supported by National Science Foundation through the
Astrophysics and Astronomy Grant Program (grants 1615100 \& 1615940),
the University of Colorado through the George Ellery Hale
visiting faculty appointment, and the Shota Rustaveli National 
Science Foundation (Georgia) (grant FR/18-1462).
Simulations presented in this work have been performed with computing resources
provided by the Swedish National Allocations Committee at the Center for
Parallel Computers at the Royal Institute of Technology in Stockholm.
This work utilised the Summit supercomputer, which is supported by the National
Science Foundation (award No.\ CNS-0821794), the University of Colorado
Boulder, the University of Colorado Denver, and the National Center for
Atmospheric Research.

\markboth{\rm ROPER POL ET AL.}{\rm GEOPHYSICAL \& ASTROPHYSICAL FLUID DYNAMICS}

\markboth{\rm ROPER POL ET AL.}{\rm GEOPHYSICAL \& ASTROPHYSICAL FLUID DYNAMICS}

\appendix

\section{Normalised GW equation}
\label{normalised_eqn}

We present here the detailed derivation of the normalised
GW \Eq{normalised_GW}.
The spatial components of the background metric for an isotropic
and homogeneous universe are described by the
Friedmann--Lema\^itre--Robertson--Walker (FLRW) metric $\gamma_{ij} (t) = a^2 (t) \delta_{ij}$,
for signature $(- + +\,\,+)$, where $a(t)$ is the scale factor, and $\delta_{ij}$ is the
Kronecker delta. Assuming small tensor-mode perturbations $a^2 (t) h_{ij}^{\rm phys} (\xx, t)$,
where $h_{ij}^{\rm phys} (\xx, t)$ are also called strains, over the
FLRW metric, such that the resulting metric tensor is $g_{ij} (\xx, t) = \gamma_{ij} (t) + a^2 (t)
h_{ij}^{\rm phys} (\xx, t) =  a^2 (t) [
\delta_{ij} + h_{ij}^{\rm phys} (\xx, t) ]$, the GW equation in physical space
and time coordinates, governing the physical strains is
\begin{equation}
\biggl( {\upartial^2\over \upartial t_{\rm phys}^2} + 3 H (t) {\upartial \over 
\upartial t_{\rm phys}} - c^2 \nabla^2_{\text{phys}}
\biggr) h_{ij, \text{phys}}^{\rm TT} (\xx, t)\, = \,\frac{16 \pi G}{c^2}
T^{\text{TT}}_{ij, \text{phys}} (\xx, t)\,,
\label{GW1}
\end{equation}
where $c$ is the speed of light, $G$ is Newton's gravitational constant, $H(t) = \dot{a}/a =
a'/a^2$ is the Hubble rate, where dot represents physical time derivative, and prime denotes
conformal time derivative; and TT is the
transverse and traceless projection described in \Sec{TT}. The physical space
and time coordinates are related to the comoving space and conformal time coordinates
through $\xx_{\rm phys} = a(t) \xx$, $a (t) {\bm \nabla}_{\text{phys}} = {\bm \nabla}$,
and ${\mathrm d}t_{\text{phys}} = a (t) {\mathrm d}t$. The physical
stress-energy tensor is related to the comoving stress-energy tensor through
$a^4(t) T_{ij, \text{phys}}^{\text{TT}} (\xx, t) = T_{ij}^{\text{TT}} (\xx, t)$.
The comoving variables are defined to take into account the expansion of the universe,
and the corresponding dilution of the energy density.
Note that the physical wavevector in Fourier space, $\kk_{\rm phys}$,
is also related to the comoving wavevector as $a(t) \kk_{\rm phys} = \kk$.
This also affects the frequency $f_{\rm phys}$, which is related 
to $k_{\rm phys} = |\kk_{\rm phys}|$
through the dispersion relation $2 \pi f_{\rm phys} = c k_{\rm phys}$. 
Hence, the comoving frequency is $a(t) f_{\rm phys} = f$.

The stress-energy has been defined as a comoving variable. This affects the hydromagnetic
fields that are part of the stress-energy tensor, i.e., $p, \rho, \uu, \BB$, and the fields
that can be obtained as a function of the hydromagnetic fields, e.g., $\JJ$ or $\pmb{S}$,
which appear in \Sec{hydromagnetic}. Hence, all these fields are defined also as comoving
fields, consistently with the $a^{-4} (t)$ dilution of the stress-energy tensor. This is
described in more detail in \cite{Sub16}.

\subsection{Using conformal time}

The GW \Eq{GW1} can be expressed in terms of comoving space and conformal
time coordinates,
comoving stress-energy tensor, and scaled strains $h_{ij}^{\text{TT}} (\xx, t) =
a (t) h_{ij, \text{phys}}^{\text{TT}} (\xx, t)$ as
\begin{equation}
\left( {\upartial^2 \over \upartial t^2} - {a'' \over a} (t) - c^2 \nabla^2 \right)
h_{ij}^{\text{TT}} (\xx, t) \,= \,\frac{16 \pi G}{a (t) c^2} T_{ij}^{\text{TT}} (\xx, t)\,.
\label{GW2}
\end{equation}

When the universe was dominated by radiation (approximately until the epoch of recombination),
the pressure was dominated by radiation, and given by the relativistic equation of state
$p = \rho c^2/3$, so the speed of sound is $\cs = c/\sqrt{3}$.
This relation allows us to obtain the solution to the Friedmann equations \citep[see][]{Friedmann1922},
which results in a linear evolution of the scale factor with conformal time, i.e., $a(t) = C_t t$,
for some constant $C_t$, such that $a'' = 0$.
This simplifies equation (\ref{GW2}) to
\begin{equation}
\left( {\upartial^2 \over \upartial t^2} - c^2 \nabla^2 \right)
h_{ij}^{\text{TT}} (\xx, t)\,= \,\frac{16 \pi G}{a (t) c^2} T_{ij}^{\text{TT}} (\xx, t)\,.
\label{GW3}
\end{equation}

The stress-energy tensor can be normalised as
$\bar{T}_{ij}^{\text{TT}} (\xx, t) = T_{ij}^{\text{TT}} (\xx, t)/
\EErad^*$, where the radiation energy
density during the radiation-dominated epoch is equal
to the total energy density of the universe, given
by the critical energy density derived from the Friedmann equations,
\begin{equation}
\EErad^* \,=\, \frac{\pi^2 g_* (T_*) (k_{\rm B} T_*)^4}{30 \, (\hbar c)^3} \,=\,
\EEcrit^* \,=\, \rho_* c^2 \,=\, \frac{3 H_*^2 c^2}{8 \pi G}\,,
\label{rho_rad}
\end{equation}
where $T_*$, $g_* (T_*)$, and $H_\ast$ are the temperature, the number
of relativistic degrees of freedom, and the Hubble rate, respectively,
during the time of generation, and $k_{\rm B}$ and $\hbar$ are the Boltzmann
and the reduced Planck constants.
In our simulations, we choose the electroweak phase transition as the
starting time
of turbulence generation. The Hubble parameter at the electroweak phase
transition is given by (\ref{rho_rad}) as
\begin{equation}
H_* \,\approx \,2.066 \cdot 10^{10} \ {\rm s}^{-1} \left(T_* \over 100 \ {\rm GeV} \right)^2
\left( g_* (T_*) \over 100 \right)^{1/2},
\label{H*}
\end{equation}
where an energy of $k_{\rm B} T_* = 100\GeV$,
and the corresponding relativistic degrees of freedom
$g_*(T_*) = 100$, have been used to get an estimate
of the Hubble rate, $H_*$, at the electroweak phase transition. The exact temperature at the 
electroweak phase transition is
uncertain, so it is left as a parameter in (\ref{H*}).

\subsection{Case I: Static universe}

The resulting expression for GWs in a static universe, after
normalisation of $\bar{T}_{ij}^{\rm TT} (\xx, t)$, is
\begin{equation}
\left( {\upartial^2\over \upartial t^2 }- c^2 \nabla^2 \right) h_{ij}^{\rm TT} (\xx, t)
\,= \,6 H_\ast^2 \bar{T}_{ij}^{\rm TT} (\xx, t)\,.
\label{GW_non_expanding}
\end{equation}
This is the equation that we use for a non-expanding universe,
where we can shift the time coordinate to zero at the starting time
of generation due to the absence of the scale factor $a(t)$ in the equation,
to simplify the resulting expressions.
Comoving space and conformal time coordinates are the same as the physical
coordinates in this case, and likewise for physical and scaled
strains. Hence, the solution is readily given
in terms of physical quantities.

\subsection{Case II: Expanding universe}

For more realistic cosmological computations of GWs
generated by hydromagnetic turbulence during
the radiation-dominated epoch, the expansion of the universe has to be taken
into account. For this purpose, it is useful to normalise $t$
with respect to the conformal time of generation $t_*$, which corresponds to the
time where the turbulent motions sourcing GWs are assumed to begin
to be generated, such that $\bar{t} = t/t_*$.
The Hubble parameter at the starting time of generation $H_*$ can be related to
$C_t$ via $C_t = H_* a_*^2$, which allows us to express the initial time of
generation as $t_* = \left( H_* a_* \right)^{-1}$. The Laplacian operator is
normalised as $\bar\nabla^2 = c^2 \nabla^2/H_*^2$.
As previously, the normalisation
of the Laplacian operator implies the
normalisation of wavenumber $\bar{k} = c k/H_*$, and the frequency $\bar{f} =
f/H_*$, related to $\bar{k}$, through the dispersion relation $2 \pi \bar{f} = \bar{k}$.
This normalisation allows us to express the evolution of the scale factor as $a (\bar{t}) =$
$a_* \bar{t}$, where $a_*$ is the scale factor at the starting time of generation.
For a flat FLRW model, the scale factor $a(t)$ at any time $t$ is only defined
relative to its value at any other time of the universe history \citep{Wei08}.
The usual convention is to set $a_0 = 1$ at the present time.
However, for the present purpose, it is more convenient to set $a_* = 1$,
at the initial time of the simulation $\bar{t} = 1$, i.e., at the
electroweak phase transition.
This needs to be taken into account when the characteristic
amplitude and the energy density obtained
from the calculations are expressed as observables at the present time,
as described in \App{spectral}.
The expansion of the universe is assumed to be adiabatic, such that
$g_S (T) T^3 a^3 (T)$ is constant, where $g_S (T)$ is the number
of adiabatic degrees of freedom at temperature $T$. This allows us to
compute the scale factor at the present time $a_0$, relative to
$a_* = 1$, as
\begin{equation}
a_0 \,\approx\, 1.254 \cdot 10^{15} \left(T_* \over 100 \ {\rm GeV}\right) 
\left(g_S (T_*) \over 100\right)^{1/3},
\label{a0}
\end{equation}
where $g_S(T_*)$ is the number of adiabatic degrees of freedom during the period
of generation.
Again, the exact values of $g_S (T_*)$ and $T_*$ at the electroweak phase transition 
are uncertain, and the previous
expression allows to use different values.
After this normalisation, the GW \Eq{GW3} reduces to the
normalised GW equation.
\begin{equation}
\left({\upartial^2 \over \upartial \bar{t}^2} - \bar\nabla^2\right)
h_{ij}^{\rm TT} (\bar{\xx}, \bar{t})\, =\, {6 \over \bar{t}} \bar{T}_{ij}^{\rm TT}
(\bar{\xx}, \bar{t})\,,
\label{GW4}
\end{equation}
where $\bar{\xx}$ are also normalised comoving space coordinates, related
to comoving space coordinates $\bar{\xx} = H_* \xx/c$.
After removing the bars, we recover (\ref{normalised_GW}).

\section{Computation of the characteristic amplitude, the GW energy density and
the spectral functions.}
\label{spectral}

We present here a detailed description of the characteristic amplitude and
GW energy density, defined in \Sec{GW_energy_density}, how to relate these quantities to
observables at the present time, and how to compute them from the numerical
simulations. From now on,
to simplify notation, the TT projection will be assumed
since only TT projected strains, i.e., the gauge invariant strains,
are physically relevant.
We also define the spectral functions $S_h (k, t)$, $S_{\dot{h}} (k, t)$,
and $A_{\dot{h}}$, that are used in \Sec{GW_energy_density} to describe the
spectral characteristic amplitude $\hrms(k, t)$, the GW energy
spectrum $\varOmega_{\rm GW} (k, t)$, and the helical GW spectrum
$\varXi_{\rm GW} (k, t)$.
The characteristic amplitude $\hrms (t)$, defined in (\ref{hrms_def}) in terms
of the physical strains, can also be expressed in terms of
the scaled strains and normalised conformal time as
\begin{equation}
\hrms^2 (t) \,=\, \frac{1}{2 t^2} \bra{h_{ij} (\xx, t) h_{ij} (\xx, t)}\,,
\end{equation}
where angle brackets indicate volume averaging in physical space.
This expression can be computed directly from the solutions that we obtain
from the code, following the methodology presented in \Sec{two_approaches}.
In our simulations, we assume MHD turbulence sourcing GWs from the time of
generation until the end of the radiation-dominated epoch, at recombination.
At this point, we assume that there are no sources affecting the primordial
stochastic GW background, such that the latter freely propagates from
this time on.
In the absence of GW generation, the physical
strains damp as $a^{-1} (t)$
with the expansion of the universe, as can be inferred from (\ref{normalised_GW})
with $T_{ij} = 0$.
The observable characteristic amplitude at
the present time, $\hrms^0$, is related to
the final result obtained at the end of the simulation
(denoted by superscript ``end'') as
\begin{equation}
\hrms^0 \,= \,\hrms^{\rm(end)} a^{\rm(end)} a_0^{-1}\, =\, \sqrt{\half \bra{h_{ij} (\xx, t )
h_{ij} (\xx, t)}^{\rm(end)}}\, a_0^{-1}\,,
\label{hrms0}
\end{equation}
where $a_0$ is the scale factor at the present time, scaled such that
the scale factor at the starting time of generation is
unity; see (\ref{a0}), and $a^{\rm (end)}$ is the scale factor
at the end of the simulation.
We have used the fact that the scale factor is $a(t) = t$ in
normalised units (see appendix~\ref{normalised_eqn} for details), to simplify the previous
equation.

The mean GW energy density $\EEGW(t)$ is defined in (\ref{EEGW_def}) in terms of the
derivative of the physical strains with respect to physical time $\dot{h}_{ij}^{\rm phys} (\xx, t)$ 
and the normalised GW energy density is
$\varOmega_{\rm GW} (t) = \EEGW(t)\left/ \EErad^*\right.$, where the radiation energy
density at the starting time of generation, $\EErad^*$, is defined in (\ref{rho_rad}).

In the case of an expanding universe, it is also useful to define normalised
$\varOmega_{\rm GW}^0 (t) = \left. \EEGW (t) \right/ \EEcrit^0$, where
$\EEcrit^0 = 3 H_0^2 c^2 \left/ \left(8 \pi G\right) \right.$
is the critical energy density at present time, $H_0$ being the current Hubble parameter.
The value of the Hubble parameter at the present time is customarily expressed as
$H_0 = 100 \, h_0 \km\s^{-1}\Mpc^{-1}$, where $h_0 \approx 0.71$ takes into account the
uncertainties of its value \citep{Riess11, Planck15}.
To get rid of the uncertainties of $H_0$ in the calculations, it is common to
use $h_0^2 \varOmega_{\rm GW}^0 (t)$, instead of just $\varOmega_{\rm GW}^0 (t)$
\citep{Mag00}.
This is used to compute the present time
observable GW signal generated during the early universe turbulent epoch and
taking into account its dilution due to the expansion of the
universe, as it has been done for $\hrms (t)$.

We recall that the dots in (\ref{EEGW_def})
denote physical time derivatives.
In terms of the normalised conformal time $t$ and scaled strains $h_{ij} (\xx, t)$,
the physical time derivative of the physical strains
$h_{ij}^{\rm phys} (\xx, t)$ is
\begin{equation}
\dot{h}_{ij}^{\rm phys} (\xx, t) \,= \,{H_*\over t}
{\upartial\over\upartial t} \left(h_{ij} (\xx, t)\over t \right)\, =\,
\frac{H_*}{t^2} \left( h'_{ij} (\xx, t) - {h_{ij} (\xx, t) \over t} \right)\,.
\label{rel_dot_prime}
\end{equation}
Therefore, the normalised GW energy density $\varOmega_{\rm GW}^0 (t)$
can be expressed in terms of the normalised conformal time as
\begin{equation}
\varOmega_{\rm GW}^0(t)\, =\, \frac{\left(H_*\left/H_0\right.\right)^2}{12 t^4}
\left[ \bra{h'_{ij}  h'_{ij}} + \frac{1}{t^2}
\bra{h_{ij}  h_{ij}} - \frac{2}{t} \bra{h'_{ij} h_{ij}} \right],
\label{Omega_conf}
\end{equation}
where the explicit dependence on $\xx$ and $t$ has been omitted
to simplify the notation.
The energy density $\varOmega^0_{\rm GW}(t)$ has three different contributions:
\begin{subequations}
\begin{align}
\varOmega_{\rm GW}^{{\rm h'},0} (t) \,\propto \,&\,\bra{h'_{ij} h'_{ij}}\,,\\
\varOmega_{\rm GW}^{{\rm h}, 0} (t) \,\propto\,&\, \bra{h_{ij}  h_{ij}}\, =\, 2 t^2 \hrms^2(t)\,,\\
\varOmega_{\rm GW}^{{\rm mix}, 0} (t)\,\propto\,&\, \bra{h'_{ij} h_{ij}}\,,
\end{align}
\end{subequations}
such that $\varOmega^0_{\rm GW}(t) = \varOmega_{\rm GW}^{{\rm h'},0} (t) +
t^{-2} \varOmega_{\rm GW}^{{\rm h}, 0} (t) - 2 t^{-1}
\varOmega_{\rm GW}^{{\rm mix}, 0} (t)$.
Note that $\varOmega_{\rm GW}(t)$ is obtained from (\ref{Omega_conf}) without the
prefactor $H_*\left/H_0\right.$, as well as $\varOmega^i_{\rm GW}(t)$, for
$i$ = h$'$, h, mix.
The energy density dilutes as $a^{-4}$ due to the expansion
of the universe. We can relate the GW energy density at the present time with
the final result obtained at the end of the simulation as
\begin{equation}
\varOmega_{\rm GW}^0 (t_{\rm end})
t_{\rm end}^4 a_0^{-4} \,=\, \varOmega_{\rm GW}^{{\rm h'}, 0} (t_{\rm end}) +
t_{\rm end}^{-2} \varOmega_{\rm GW}^{{\rm h}, 0} (t_{\rm end}) -
2t_{\rm end}^{-1} \varOmega_{\rm GW}^{{\rm mix}, 0} (t_{\rm end})\,,
\label{varOmegaGW0}
\end{equation}
where
\begin{subequations}
\begin{align}
\varOmega_{\rm GW}^{{\rm h'},0} (t_{\rm end}) \,= \,&\,
{\left(H_*\left/H_0\right.\right)^2
\over 12 a_0^4} \bra{h'_{ij} h'_{ij}}^{\rm(end)\,}, \\
\varOmega_{\rm GW}^{{\rm h},0} (t_{\rm end}) \,= \,&\,
{\left(H_*\left/H_0\right.\right)^2
\over 12 a_0^4} \bra{h_{ij} h_{ij}}^{\rm(end)}\,, \\
\varOmega_{\rm GW}^{{\rm mix},0} (t_{\rm end})\, = \,&\,
{\left(H_*\left/H_0\right.\right)^2
\over 12 a_0^4} \bra{h_{ij} h'_{ij}}^{\rm(end)}\,.
\end{align}
\end{subequations}

The spectral function $S_h (k, t)$, used to compute the characteristic
amplitude $\hrms(k, t)$ (see \Sec{GW_energy_density}) is defined as
\begin{equation}
S_h (k, t) \,= \,\int_{\varOmega_D}
\left(\left|\tilde{h}_+^{\rm phys}(\kk,t)\right|^2+
\left|\tilde{h}_\times^{\rm phys}(\kk,t)\right|^2\right) \,
k^{D - 1} \, \dd \varOmega_k\,.
\label{Sh_app}
\end{equation}
This is the shell-integrated spectrum of $\half \tilde{h}^{\rm phys}_{ij}
\tilde{h}^{\rm phys, *}_{ij}$ over all directions of $\kk$, which
can be expressed, using the orthogonality property of the linear polarisation
basis, as $\tilde{h}^{\rm phys}_+ \tilde{h}_+^{\rm phys, *} + 
\tilde{h}^{\rm phys}_\times \tilde{h}_\times^{\rm phys, *}$; 
see (\ref{orthogonality});  $D$ is the number of dimensions, and
$\varOmega_D$ is the solid angle subtended by the entire $(D - 1)$--sphere, 
such that $\varOmega_1 = 2$, $\varOmega_2 =
2 \pi$, $\varOmega_3 = 4\pi$ are, respectively, the $(D - 1)$--surface of a line, a
circle and a sphere.
The spectral function $S_h (k, t)$ is expressed in terms of the scaled
strains $\tilde{h}_{+,\times} (\kk, t)$ as
\begin{equation}
S_h(k, t)\, =\, {1\over t^2} \int_{\varOmega_D}
\left(\left|\tilde{h}_+(\kk,t)\right|^2+\left|\tilde{h}_\times(\kk,t)
\right|^2\right) \, k^{D - 1} \, \dd \varOmega_k\,.
\label{Sh_app2}
\end{equation}
For a non-expanding universe we can directly use (\ref{Sh_app})
since scaled and physical strains are then the same.
The integration over all wavenumbers of $S_h(k, t)$ leads to
\begin{equation}
\int_0^\infty S_h(k, t) \, \dd k\, = \,\int_{-\infty}^\infty h_{\rm rms}^2
(k, t) \, \dd \ln k = h_{\rm rms}^2 (t)\,.
\end{equation}
This can be shown making use of the Parseval's theorem
\begin{align}
\bra{h_{ij} (\xx, t) h_{ij} (\xx, t)}\, =\,&\,
\frac{1}{L^3} \int_{\cal V} h_{ij} (\xx, t) h_{ij} (\xx, t) \, \dd^3 \blue{\xx} \nonumber \\ 
=\,&\, \,
\sum_{\kk} \tilde{h}_{ij} (\kk, t) \tilde{h}_{ij}^\ast (\kk, t)
\nonumber \\
=\,&\, \,2 \sum_{\kk} \left( \tilde{h}_+ (\kk, t)
\tilde{h}_+^* (\kk, t) + \tilde{h}_\times (\kk, t)
\tilde{h}_\times^* (\kk, t) \right)\,\nonumber \\
= \,&\,2 t^2 \int_0^{\infty}
S_h (k, t) \, \dd k \,= \,2 t^2 \hrms^2(t)\,,
\label{Parseval}
\end{align}
where ${\cal V}=L^3$ is the volume in physical space of length $L$.

The spectral function associated with the characteristic amplitude
$h^0_{\rm rms} (k)$ can also be expressed as an observable at the present
time, as it has been done in (\ref{hrms0}), for the physical $\hrms(t)$.
\begin{equation}
h^0_{\rm rms} (k)\, =\, a_0^{-1}\sqrt{k S_h (k, t_{\rm end})}\,,
\end{equation}
where $t_{\rm end}$ is the end time of the simulation.
When we compute the observable
$\hrms$ spectrum at the present time, we are interested in its dependence
on the physical wavenumber $k^{\rm phys}_0$ at the present time.
The relation to the normalised wavenumber $k$ is given by
\begin{equation}
k_0^{\rm phys} \,= \,H_* a_0^{-1} k/c\,.
\label{kshift}
\end{equation}
This shifting in wavenumbers is computed for the following spectral functions
when we plot spectra as observables at the present time.

Analogous to $S_h (k, t)$ we define the spectral function $S_{\dot{h}} (k, t)$,
with a dot, as
\begin{equation}
S_{\dot{h}}(k, t)\, = \,\int_{\varOmega_D}\left( \left|\dot{\tilde{h}}_{+}^{\rm phys}
(\kk, t) \right|^2 + \left|\dot{\tilde{h}}_\times^{\rm phys} (\kk, t)
\right|^2 \right) \,k^{D - 1}\, \dd\varOmega_k\,.
\label{Sdoth_app}
\end{equation}
This is the shell-integrated spectrum over all directions of
$\half \dot{\tilde{h}}_{ij} ^{\rm phys} \dot{\tilde{h}}_{ij}^{\rm phys, *} =
\dot{\tilde{h}}_+^{\rm phys} \dot{\tilde{h}}_+^{\rm phys, *} +
\dot{\tilde{h}}_\times^{\rm phys} \dot{\tilde{h}}_\times^{\rm phys, *}$,
defined as in (\ref{Sh_app}).
The corresponding spectral function for the energy spectrum
$E_{\rm GW}(k,t)$ is defined as
\begin{equation}
E_{\rm GW} (k, t) \,= \,\frac{c^2}{16 \pi G} S_{\dot{h}} (k, t),\hskip 8mm
\text{such that \ }\hskip 6mm \int_0^{\infty} E_{\rm GW} (k, t)\, \dd k\, =\,
\EEGW (t)\,,\hskip 6mm
\label{EGW_app}
\end{equation}
where, as before, the resulting mean energy density $\EEGW (t)$ corresponds to that
defined in (\ref{EEGW_def}), due to Parseval's theorem, used in (\ref{Parseval}).
The GW spectrum $E_{\rm GW} (k, t)$ is used to define a characteristic
length scale of GWs, $\xi_{\rm GW} (t)$, and the corresponding characteristic
wavenumber, $k_{\rm GW} (t)$, as
\begin{equation}
\xi_{\rm GW} (t) \,=\, k_{\rm GW}^{-1} (t)\, =\, \frac{1}{\EEGW (t)}
\int_0^{\infty}k^{-1}  E_{\rm GW} (k, t) \, \dd k\,.
\label{kGW1}
\end{equation}

Note that we have defined here $\xi_{\rm GW}$ without $2\pi$ factor, which
is analogous to our definition of the magnetic and kinetic correlation
lengths \citep[see][]{KBTR10}.
We also define $\varOmega_{\rm GW}^0
(k, t)$, normalising with $H_0$ instead of $H_*$.
The antisymmetric spectral function of GWs, $A_{\dot{h}} (k, t)$, in relation to the
symmetric spectral function $S_{\dot{h}} (k, t)$, is defined as
\begin{align}
\blue{i} A_{\dot{h}} (k, t)\, &\, = \,\int_{\varOmega_D}\left(
\dot{\tilde{h}}_{+}^{\rm phys}
\dot{\tilde{h}}_{\times}^{{\rm phys}, \ast} -
\dot{\tilde{h}}_{+}^{{\rm phys},\ast}
\dot{\tilde{h}}_{\times}^{\rm phys} \right) \, k^{D - 1} \, \dd \varOmega_k
\nonumber \\
&\, =\, 2\blue{i} \int_{\varOmega_D} {\rm Im} \left( \dot{\tilde{h}}_{+}^{\rm phys}
\dot{\tilde{h}}_{\times}^{{\rm phys}, \ast} \right)
\, k^{D - 1} \, \dd  \varOmega_k\,,
\label{Adoth_app}
\end{align}
where the explicit dependence on $\kk$ and $t$ in the integrand has been
avoided for notational simplicity.
Contrary to the symmetric spectral function $S_{\dot{h}} (k, t)$, which is
positive definite, $A_{\dot{h}} (k, t)$ can be positive or negative.
The motivation to define $S_{\dot{h}}(k, t)$ and $A_{\dot{h}} (k, t)$
follows the description of the autocorrelation function for any second-rank
tensor used in \cite{CDK04}. The autocorrelation function for the
physical time derivative of the scaled strains $\dot{\tilde{h}}_{ij}(\kk, t)$
is defined in terms of the symmetric and antisymmetric
spectral functions as
\begin{equation}
\int_{\varOmega_D}
\dot{\tilde{h}}_{ij} (\kk, t) \dot{\tilde{h}}^*_{lm} (\kk', t)
 \,k^{D - 1}\, \dd\varOmega_k\, =\,
\frac{1}{4} \Big[ {\cal M}_{ijlm} S_{\dot{h}} (k, t) + i {\cal A}_{ijlm}
A_{\dot{h}}(k, t) \Big]\,,
\end{equation}
where the prefactor tensors ${\cal M}_{ijlm}$, and ${\cal A}_{ijlm}$ are
defined as
\begin{align}
{\cal M}_{ijlm}\, =\, &\,P_{il} P_{jm} + P_{im} P_{jl} - P_{ij} P_{lm}\,,\\
{\cal A}_{ijlm} \,=\,&\, \half \ee^q \left(P_{jm} \epsilon_{ilq} + P_{il}
\epsilon_{jmq} + P_{im} \epsilon_{jlq} + P_{jl} \epsilon_{imq} \right)\,,
\end{align}
where $P_{ij}$ is the projection operator that appears in (\ref{Tij_TT}),
$\ee^i$ is the basis defined in \Sec{SignSwap}, and $\epsilon_{ijk}$
is the Levi-Civita tensor.
The spectral function associated with the corresponding antisymmetric
contribution to the energy density,
$H_{\rm GW} (k, t)$, is
\begin{equation}
H_{\rm GW}(k, t)\, = \,\frac{c^2}{16 \pi G} A_{\dot{h}} (k, t), \hskip 8mm \text{such that \ }\hskip 6mm
\int_0^{\infty} H(k, t)\, \dd k\, =\, {\cal H}_{\rm GW} (t)\,,\hskip 6mm
\label{HGW_app}
\end{equation}
where an antisymmetric or helical energy density ${\cal H}_{\rm GW} (t)$ has been defined
in analogy to the energy density ${\cal E}_{\rm GW} (t)$.
We also define $\varXi_{\rm GW}^0 (k, t)$ and
$\varXi_{\rm GW}^0 (t)$ normalised with $H_0$.

The spectral functions $S_{\dot{h}} (k, t)$ and $A_{\dot{h}} (k, t)$
depend on the physical time derivatives of the physical strains.
We want to express the spectral functions in terms
of derivatives of the scaled strains with respect to conformal time.
For that purpose, to compute $S_{\dot{h}} (k, t)$, we define
the spectral functions $S_{h'} (k, t)$ and $S_{\rm mix}(k, t)$ as
\begin{align}
S_{h'}(k,t)\, = \,&\, \int_{\varOmega_D}
\left(\left|\tilde{h}'_+(\kk,t)\right|^2+\left|\tilde{h}'_\times(\kk,t)
\right|^2\right) \, k^{D - 1} \, \dd \varOmega_k\,, \\
S_{\rm mix}(k,t)\, =\,&\, \int_{\varOmega_D} {\rm Re}
\left( \tilde{h}_+ \tilde{h}_+^{'*} (\kk, t) +
\tilde{h}_\times \tilde{h}_\times^{'*} (\kk, t)
\right) \, k^{D - 1} \, \dd \varOmega_k\,.
\end{align}
Now, using (\ref{rel_dot_prime}), we can express $S_{\dot{h}}(k, t)$ as
\begin{equation}
S_{\dot{h}} (k, t) \,=\, \frac{H_*^2}{t^4} \Big[ S_{h'}(k, t) +
S_h (k, t) - 2 t^{-1} S_{\rm mix} (k, t)\Big]\,,
\label{Shhprime}
\end{equation}
where $S_h(k, t)$ is defined in (\ref{Sh_app}) and (\ref{Sh_app2}).
This allows us to define, analogous to (\ref{EGW_app}) and (\ref{OmegaGW_app}),
the GW energy spectra
\begin{equation}
E_{\rm GW}^{\rm h'} (k, t)\, =\, c^2 S_{h'} (k, t)\left/ \left(16 \pi G \right) \right., \hskip 8mm
E_{\rm GW}^{\rm mix} (k, t)\, =\, c^2 S_{\rm mix} (k, t)\left/ \left(16 \pi G \right) \right.,\hskip8mm
\end{equation}
and the normalised spectra
\begin{subequations}
 \label{Omega_GW_h'_app}
\begin{align}
\varOmega_{\rm GW}^{\rm h'} (k, t)\, =\,&\, k S_{h'} (k, t) \left/
\left( 6H_*^2 \right) \right., &
\varOmega_{\rm GW}^{\rm mix} (k, t) \,= \,&\,k S_{\rm mix} (k, t) \left/
\left( 6H_*^2 \right) \right., \\
\varOmega_{\rm GW}^{{\rm h'}, 0} (k, t)\, = \,&\,k S_{h'} (k, t) \left/
\left( 6H_0^2 \right) \right., &
\varOmega_{\rm GW}^{{\rm mix}, 0} (k, t)\, = \,&\,k S_{\rm mix} (k, t) \left/
\left( 6H_0^2 \right) \right..\hskip 8mm
\end{align}
\end{subequations}

For the antisymmetric spectral function we define $A_{h'} (k, t)$,
$A_{\rm mix} (k, t)$ and $A_h (k, t)$ as
\begin{align}
A_{h'} (k, t)\, = \,&\,2 \int_{\varOmega_D} {\rm Im} \left( \tilde{h}'_+
\tilde{h}_\times^{\prime \, \ast} \right)
\, k^{D - 1} \, \dd \varOmega_k\, , \\
A_{\rm mix} (k, t)\, =\,&\, \int_{\varOmega_D} {\rm Im} \left( \tilde{h}'_+
\tilde{h}_\times^{\ast} + \tilde{h}_+ \tilde{h}^{\prime \, \ast}_\times \right)
\, k^{D - 1} \, \dd\varOmega_k\, , \\
A_h (k, t) \,= \,&\,2 \int_{\varOmega_D} {\rm Im} \left( \tilde{h}_+^{\rm phys}
\tilde{h}_\times^{\rm phys, \ast} \right)
\, k^{D - 1} \, \dd\varOmega_k \nonumber\\
= \,&\,2 t^{-2} \int_{\varOmega_D} {\rm Im}
\left( \tilde{h}_+ \tilde{h}_\times^{\ast} \right)
\, k^{D - 1} \, \dd\varOmega_k\, ,
\end{align}
such that $A_{\dot{h}} (k, t)$ can be expressed as
\begin{equation}
A_{\dot{h}} (k, t) \,=\, \frac{H_*^2}{t^4} \Big[ A_{h'} (k, t) + A_h (k, t) -
2 t^{-1} A_{\rm mix} (k, t)\Big]\,.
\end{equation}
Again, this allows us to define, analogous to (\ref{HGW_app}) and (\ref{XiGW_app}),
the antisymmetric GW energy spectra
\EQ
H_{\rm GW}^{\rm h'} (k, t)\, = \,c^2 A_{h'} (k, t)\left/ \left(16 \pi G \right) \right., \hskip 8mm
H_{\rm GW}^{\rm mix} (k, t)\, =\, c^2 A_{\rm mix} (k, t)\left/ \left(16 \pi G \right) \right.,\hskip 8mm
\EN
and the normalised spectra
\begin{subequations}
\begin{align}
\varXi_{\rm GW}^{\rm h'} (k, t)\, = \,&\,k A_{h'} (k, t) \left/
\left( 6H_*^2 \right) \right., &
\varXi_{\rm GW}^{\rm mix} (k, t) \,=\,&\, k A_{\rm mix} (k, t) \left/
\left( 6H_*^2 \right) \right., \\
\varXi_{\rm GW}^{{\rm h'}, 0} (k, t)\, =\,&\, k A_{h'} (k, t) \left/
\left( 6H_0^2 \right) \right., &
\varXi_{\rm GW}^{{\rm mix}, 0} (k, t)\, =\,&\, k A_{\rm mix} (k, t) \left/
\left( 6H_0^2 \right) \right..\hskip 6mm
\end{align}
\end{subequations}
Therefore, the spectral functions $\varOmega_{\rm GW}^0 (k, t)$,
$\varXi_{\rm GW}^0 (k, t)$, and ${\cal P} (k, t)$ can be expressed as
\begin{align}
\varOmega^0_{\rm GW} (k, t)\, = \,&\,{\left(H_* \left/ H_0 \right. \right)^2
\over 6t^4} k \Big[ S_{h'} (k, t) + S_h (k, t) - 2 t^{-1} S_{\rm mix}
(k, t) \Big]\,, \label{varOmegaGW02} \\
\varXi^0_{\rm GW} (k, t) = \,&\,{\left(H_* \left/ H_0 \right. \right)^2
\over 6t^4} k \Big[ A_{h'} (k, t) + A_h (k, t) - 2 t^{-1} A_{\rm mix}
(k, t) \Big]\,, \label{XiGW02} \\
{\cal P} (k, t) \,= \,&\,\frac{A_{h'} (k, t) + A_h (k, t) - 2 t^{-1} A_{\rm mix} (k, t)}
{S_{h'} (k, t) + S_h (k, t) - 2 t^{-1} S_{\rm mix} (k, t)}\,.
\end{align}
Note that $\varOmega_{\rm GW} (k, t)$ and $\varXi_{\rm GW} (k, t)$ can be obtained
using (\ref{varOmegaGW02}) and (\ref{XiGW02}) without the factor $H_* \left/ H_0 \right.$.
We are interested in expressing the spectral functions corresponding
to the energy densities as observables at the present time.
These functions are obtained in the same way as (\ref{varOmegaGW0}):
\begin{align}
  \varOmega_{\rm GW}^0 (k, \, t_{\rm end})&\, t_{\rm end}^4 a_0^{-4}\nonumber\\
  =\,&\,
{\left(H_* \left/ H_0 \right. \right)^2
\over 6 a_0^4} k \Big[ S_{h'} (k,t_{\rm end}) + S_h (k, t_{\rm end})
- 2 t^{-1} S_{\rm mix} (k, t_{\rm end}) \Big]\,, \hskip8mm\label{varOmegaGW03} \\
\varXi_{\rm GW}^0 (k, \, t_{\rm end})& \,t_{\rm end}^4 a_0^{-4}\nonumber\\
=\,&\,
{\left(H_* \left/ H_0 \right. \right)^2
\over 6 a_0^4} k \Big[ A_{h'} (k,t_{\rm end}) + A_h (k,t_{\rm end})
- 2 t^{-1} A_{\rm mix} (k,t_{\rm end}) \Big]\,.\hskip8mm
\label{XiGW03}
\end{align}
These spectra are more meaningfully described as functions of the physical wavenumber
$k_0^{\rm phys}$ at the present time, given by (\ref{kshift}).

The magnetic spectrum, $E_{\rm M} (k, t)$, is defined as
\begin{subequations}
\label{EM}
\begin{equation}
2 E_{\rm M} (k, t) \,=\, \int_{\varOmega_{\rm D}} \tilde{B}_i (\kk, t)
\tilde{B}^*_i (\kk, t) k^{{\rm D} - 1}\, \dd \varOmega_k\,,
\end{equation}
such that
\begin{equation}
\int_0^{\infty} E_{\rm M} (k, t) \, \dd k\, =\,
{\cal E}_{\rm M} (t)\,,
\end{equation}
\end{subequations}
where $\tilde{B}_i (\kk, t)$ are the Fourier transformed comoving
components of the magnetic field. Again, using Parseval's theorem,
the energy density ${\cal E}_{\rm M} (t)$ is the same as that defined
in \Sec{GW_energy_density}.
In the same way, the kinetic spectrum, $E_{\rm K} (k, t)$, is defined as
\begin{subequations}
\label{EK}
\begin{equation}
2 E_{\rm K} (k, t) \,=\, \int_{\varOmega_{\rm D}} \tilde{u}_i (\kk, t)
\tilde{u}^*_i (\kk, t) k^{{\rm D} - 1}\, \dd \varOmega_k\,,
\end{equation}
such that
\begin{equation}
\int_0^{\infty} E_{\rm K} (k, t) \, \dd k\, =\,
{\cal E}_{\rm K} (t)\,,
\end{equation}
\end{subequations}
where $\tilde{u}_i (\kk, t)$ are the Fourier transformed comoving
components of the velocity field.

The characteristic length and wavenumber corresponding to the source
(kinetic or magnetic) are defined in analogy to the integral length scale in
isotropic and homogeneous turbulence, in terms of $\varOmega_{\rm M, K} (k, t)$, as
\begin{equation}
\xi_{\rm M, K} (t) \,=\, k_{\rm M, K}^{-1} (t)\, = \,\frac{1}{\varOmega_{\rm M, K} (t)}
\int_0^{\infty} k^{-2} \varOmega_{\rm M, K} (k, t) \, \dd k\,.
\label{kM}
\end{equation}

\label{lastpage}
\end{document}